\documentclass[fleqn,usenatbib]{mnras}

\usepackage{newtxtext,newtxmath}

\usepackage[T1]{fontenc}

\DeclareRobustCommand{\VAN}[3]{#2}
\let\VANthebibliography\thebibliography
\def\thebibliography{\DeclareRobustCommand{\VAN}[3]{##3}\VANthebibliography}

\usepackage{graphicx}	
\usepackage{amsmath}	
\usepackage{newtxtext,newtxmath}
\usepackage{mathtools}
\DeclarePairedDelimiter{\ceil}{\lceil}{\rceil}
\usepackage{enumitem}
\usepackage{xcolor}
\usepackage{subcaption}

\usepackage{newtxtext,newtxmath}
\DeclareSymbolFont{CMletters}{OML}{cmm}{m}{it}
\DeclareMathSymbol{\nu}{\mathord}{CMletters}{23}
\DeclareMathSymbol{v}{\mathord}{CMletters}{`v}

\title[Multi-messenger Signatures Of SMBHBs]{Multi-messenger Time-domain Signatures Of Supermassive Black Hole Binaries}

\author[M.~Charisi et al.]{Maria Charisi,$^{1}$ \thanks{E-mail: maria.charisi@nanograv.org} Stephen~R.~Taylor,$^{1}$
Jessie Runnoe,$^{1}$ 
Tamara Bogdanovic,$^{2}$
Jonathan R. Trump$^{3}$
\\
$^{1}$Department of Physics \& Astronomy, Vanderbilt University, 2301 Vanderbilt Place, Nashville, TN 37235, USA\\
$^{2}$School of Physics \&
Center for Relativistic Astrophysics, Georgia Institute of Technology, 837 State St. NW, Atlanta GA 30332, USA\\
$^{3}$Department of Physics, University of Connecticut, 196A Auditorium Road Unit 3046, Storrs, CT 06269, USA
}

\pubyear{2021}

\begin{document}
\label{firstpage}
\pagerange{\pageref{firstpage}--\pageref{lastpage}}
\maketitle

\begin{abstract}
Supermassive black hole binaries (SMBHBs) are a natural outcome of galaxy mergers and should form frequently in galactic nuclei. Sub-parsec binaries can be identified from their bright electromagnetic emission, e.g., Active Galactic Nuclei (AGN) with Doppler shifted broad emission lines or AGN with periodic variability, as well as from the emission of strong gravitational radiation. The most massive binaries (with total mass $>10^8 M_{\odot}$) emit in the nanohertz band and are targeted by Pulsar Timing Arrays (PTAs). Here we examine the synergy between electromagnetic and gravitational wave signatures of SMBHBs. We connect both signals to the orbital dynamics of the binary and examine the common link between them, laying the foundation for joint multi-messenger observations. We find that periodic variability arising from relativistic Doppler boost is the most promising electromagnetic signature to connect with GWs. We delineate the parameter space (binary total mass/chirp mass versus binary period/GW frequency) for which joint observations are feasible. Currently multi-messenger detections are possible only for the most massive and nearby galaxies, limited by the sensitivity of PTAs. However, we demonstrate that as PTAs collect more data in the upcoming years, the overlapping parameter space is expected to expand significantly.
\end{abstract}




\section{Introduction}
\label{sec:intro}
Observations of nearby galaxies have established that every massive galaxy hosts a supermassive black hole (SMBH) in the center \citep{Kormendy2013}. Additionally, in hierarchical structure formation, galaxies are built-up through frequent mergers of lower-mass progenitors \citep[e.g.,][]{white78,white91}. 
As a result, supermassive black hole binaries (SMBHBs) should form frequently in galactic nuclei \citep{Haehnelt2002}. 

The initial stages of galaxy mergers have been repeatedly observed, with dozens of dual AGN (i.e., galaxies with two active SMBHs) resolved across multiple wavelengths (see \citealt{DeRosa2019} for a review). However, sub-parsec SMBHBs have escaped detection, despite their expected ubiquity. The reason for this is that at these separations, they are extremely challenging (if not impossible) to directly resolve \citep{2018ApJ...863..185D}. However, since galaxy mergers channel significant amounts of gas to the nuclear region \citep{BarnesHernquist1992,2005ApJ...620L..79S,2006ApJ...645..986R}, binaries are likely surrounded by gas. 
The gas may accrete onto the SMBHs and produce bright electromagnetic emission, allowing us to infer their presence from their dynamical interaction with the surrounding gas \citep{2021arXiv210903262B}.

There are two main methods to detect binaries at sub-parsec separations. At  separations where enough gas can be bound to one (or both) SMBH(s) to produce prominent broad emission lines, the orbital motion of the binary will appear as a Doppler shift in these lines \citep{Begelman1980}. At smaller separations (e.g., milli-parsecs) broad emission lines are truncated, and the most promising method is to detect AGN or quasars with periodic fluctuations in their photometric lightcurves.
This could be due to streams of gas from a circumbinary disk periodically accreting onto the SMBHs, or due to a Doppler boost of the emission from gas bound to each individual SMBH, which move with relativistic velocities. Systematic searches for the above signatures in spectroscopic and photometric surveys have revealed a few hundred candidates (see \citealt{DeRosa2019} for a review).

Additionally, since binaries are strong sources of low-frequency GWs, they can also be detected via their GW emission. Binaries with mass of $10^5-10^{7}M_{\odot}$ emit GWs in the millihertz band, and will be detectable in the future by the Laser Interferometer Space Antenna (LISA; \citealt{2017arXiv170200786A}), whereas binaries with masses of $10^8-10^{10}M_{\odot}$ emit nanohertz-frequency GWs and are targeted by Pulsar Timing Arrays (PTAs).

PTAs like the North American Nanohertz Observatory for Gravitatioal waves (NANOGrav; \citealt{McLaughlin2013,2019BAAS...51g.195R}), the Parkes Pulsar Timing Array (PPTA; \citealt{2013PASA...30...17M,Hobbs2013}), and the European Pulsar Timing Array (EPTA; \citealt{2013CQGra..30v4009K}) monitor dozens of stable (very precisely rotating) milli-second pulsars. If PTAs identify deviations in the times of arrival (TOAs) of radio pulses, observed across multiple pulsars in the array and spatially correlated with a characteristic quadrupolar pattern, then GWs can be detected \citep{Hellings1983}. 

PTAs are expected to detect the stochastic GW background from a population of SMBHBs within the next few years  \citep{2016ApJ...817...70T,2021ApJ...911L..34P}, and there may be early signs of this already \citep{2020ApJ...905L..34A,2021arXiv210712112G}. This detection will illuminate the massive-galaxy merger rate, the co-evolution of SMBHs with their host galaxies, and the highly uncertain orbital evolution of SMBHBs \citep{2019BAAS...51c.336T, Burke-Spolaor2019}. Soon after the detection of the background, PTAs will also detect continuous GWs from individual SMBHBs that produce signals strong enough to stand above the background \citep{Rosado2015,Mingarelli2017,Kelley2018,2020PhRvD.102h4039T}.

Since SMBHBs are expected to emit both bright electromagnetic signals and strong GWs, they make exceptional targets for multi-messenger observations \citep{2012MNRAS.420..860S,2019BAAS...51c.490K,2019BAAS...51c.123B}.
Multi-messenger observations will offer significant advantages in constraining the dynamical interaction of binaries with their environment, unveiling the physics of accretion in the presence of binaries, and probing the co-evolution of SMBHs with their host galaxies, and more \citep{2019BAAS...51c.490K}. For instance,  \citet{2020ApJ...900..102A} demonstrated that incorporating binary orbital constraints from electromagnetic observations (e.g., period and mass of the candidate) can improve GW upper limits by a factor of two.
Similarly, including priors from electromagnetic data in PTA searches can boost the detectability of binaries and improve parameter estimation \citep{2021arXiv210508087L}.

In this paper, we explore the prospects of linking together the electromagnetic and the GW signatures of SMBHBs. Since time-domain surveys and PTAs probe binaries with similar masses and periods, it is reasonable to combine the data in a joint multi-messenger dataset. This paper presents the building blocks for a joint analysis (which will follow in a companion paper), and is organized as follows: In \S~\ref{sec:orbits}, we present the two-body problem for circular and elliptical orbits. In \S~\ref{sec:GWs} and \S~\ref{sec:em_signals}, we
connect the GW and electromagnetic signatures to the binary orbit. We also summarize how these signatures have been used for SMBHB searches and the current status of binary candidates or GW upper limits. 
In \S~\ref{sec:mma}, we connect the two signals and delineate the parameter space, in which joint multi-messenger observations are possible given the current and projected sensitivity of PTAs and time-domain surveys. In \S~\ref{sec:discussion}, we discuss caveats of our study, while in \S~\ref{sec:summary} we present future directions and briefly summarize our findings.

\section{Binary Orbits}
\label{sec:orbits}
We briefly derive the main equations of the two-body problem as they pertain to the dynamics of bound binaries in circular and eccentric orbits. For simplicity, and for consistency with previous studies that focused primarily on circular orbits, we will examine circular and eccentric orbits separately. 
Throughout, we model a binary with total mass $M_{\rm tot}=M_1+M_2$, where $M_1$ is the mass of the primary (more massive) SMBH and $M_2$ the mass of the secondary (less massive) SMBH. The SMBHs are orbiting at distances $R_1$, and $R_2$ from the common center of mass, with orbital velocities $v_1$ and $v_2$, respectively. 

\subsection{Circular Orbit}

In the simplest case, the binary is moving in a circular orbit, i.e., each SMBH is moving in a circular orbit around the common center of mass. The distance between the two SMBHs ($R=R_1+R_2$) is constant. The two SMBHs are always at the opposite sides of the common center of mass, orbiting with anti-parallel velocities $v_1=\omega R_1$ and $v_2=\omega R_2$, where $\omega=2\pi/P$ is the orbital angular frequency and $P$ is the orbital period. In the simplest scenario all of the orbital parameters are constant (i.e., losses to GWs are negligible). 

By equating the gravitational force to the centripetal force, we get
\begin{align} \label{eq:grav_force}
    \frac{GM_1M_2}{R^2} &=\frac{M_1v_1^2}{R_1}=M_1\omega^2R_1\\    
    &=\frac{M_2v_2^2}{R_2}=M_2\omega^2R_2.
\end{align}
Therefore, the orbital velocities of the two SMBHs and the radii of their orbits are related as 
\begin{equation}
    v_1/v_2=R_1/R_2=M_2/M_1\equiv q
\end{equation}
where $q$ is the mass ratio of the binary, with $0\leq q\leq 1$. We note that $q=1$ corresponds to an equal-mass binary, whereas the limiting case of $q=0$ corresponds to the circular motion of one body in a central potential. Additionally, since $R=R_1+R_2=R_1+M_1/M_2R_1=(M_1+M_2)/M_2R_1$, 
it is straightforward to derive Kepler's third law:
\begin{equation}\label{eq:KeplerLaw}
    R^3=\frac{G(M_1+M_2)}{4\pi^2}P^2=\frac{GM_{\rm tot}}{4\pi^2}P^2.
\end{equation}
Next, we use \autoref{eq:KeplerLaw} to express the velocities $v_1$, $v_2$ in terms of the binary parameters $M_{\rm tot}$, $q$ and $P$,
\begin{equation}\label{eq:velocities}
    v_1=\left(\frac{q}{1+q}\right)\left(\frac{2\pi GM_\mathrm{tot}}{P}\right)^{1/3}, \,\, v_2=\frac{v_1}{q}=\left(\frac{1}{1+q}\right)\left(\frac{2\pi GM_\mathrm{tot}}{P}\right)^{1/3} 
\end{equation}

\subsection{Eccentric Orbit}

For simplicity, let's first consider an eccentric binary in the limit of low mass ratio ($q\simeq0$). We return to the general case for all mass ratios at the end of the section. In this case, the primary SMBH is fixed at one focus of the ellipse, and the motion of the secondary SMBH with respect to the primary in polar coordinates is given by the conic section equation for an ellipse
\begin{equation}
\label{eq:r_vs_f}
    R=\frac{a(1-e^2)}{1+e\cos f}
\end{equation}
where $a$ is the semi-major axis, $0\leq e\leq1$ is the eccentricity (with $e=0$ corresponding to a perfectly circular orbit), and $f$ is an angle called true anomaly, measured from the periapsis of the orbit.

\autoref{eq:r_vs_f} expresses the position $R$ as a function of the angle $f$. However, unlike the circular orbit, calculating the position as a function of time for a given initial position is not trivial. This is known as the Kepler problem, the derivation of which is beyond the scope of this paper; we refer the reader to \citet{2010exop.book...15M}. In order to derive the position $R$ as a function of time, we need to take the time derivative $\dot{R}$, and eliminate the angle dependence. Then the differential equation of motion becomes
\begin{equation}
    \label{eq:R_dot_R}
    \dot{R}=\frac{2\pi a}{P R}\sqrt{a^2e^2-(R-a)^2}
\end{equation}
This equation can be solved by introducing a new variable called eccentric anomaly $E$, which is related to the position as 
\begin{equation}
\label{eq:R_vs_E}
    R= a (1-e\cos E).
\end{equation}
Then the differential \autoref{eq:R_dot_R} transforms to 
\begin{equation}
    \dot{E}=\frac{2\pi/P}{1-e\cos E}
\end{equation}
the solution of which is
\begin{equation}
\label{eq:mean_anomaly}
    l\equiv\frac{2\pi}{P}(t-t_0)=E-e\sin E
\end{equation} 
where $l$ is a new variable we introduced called mean anomaly, and $t_0$ is the time of periastron passage. 
The mean anomaly is an angle, but it does not have a geometric meaning. It is however convenient because it is a linear function of time.

For any given time $t$, from \autoref{eq:mean_anomaly}, we can determine the mean anomaly $l$ and the eccentric anomaly $E$. The latter needs to be determined numerically, and for this, we use the python package \texttt{Kepler}.\footnote{\url{ https://pypi.org/project/kepler.py/}} Then, from \autoref{eq:R_vs_E}, we can determine the position $R$ and finally from \autoref{eq:r_vs_f}, we get the true anomaly $f$.

\begin{figure}
 \includegraphics[trim={37cm 8cm 8cm 5cm}, clip, width=\columnwidth]{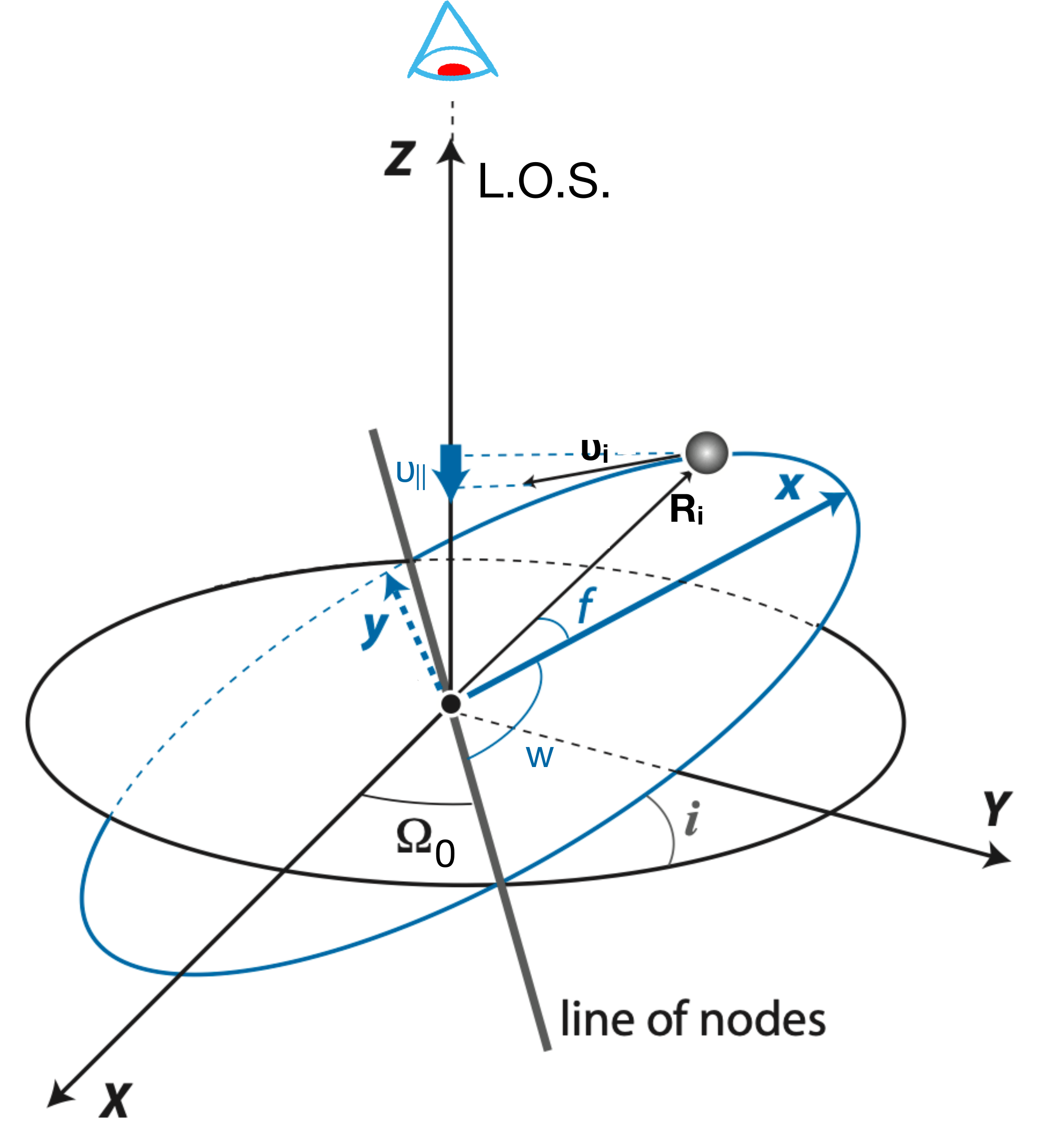}
 \caption{Orientation of an eccentric orbit with respect to the observer. The line-of-sight is aligned with the Z-axis, with the X and Y-axes defining the plane of the sky. The x and y-axes define the orbital plane, with $\hat{x}$ pointing towards the pericenter. The orbit is inclined with respect to the plane of the sky by an inclination angle $i$ and the two planes intersect at the line of nodes. The ascending node is at an angle $\Omega_0$ from the X-axis, measured in the counterclockwise direction. The argument of periapse $w$ and true anomaly $f$ are measured along the orbit from the ascending node and the periapse, respectively.}
 \label{Fig:eccentric_orbit}
\end{figure}

Next, we consider the orbit in three dimensional space and with respect to the observer. The orbital motion of the secondary is confined to a plane called the orbital plane. If we define a Cartesian coordinate system ($x,y,z$), with the $x$ and $y$-axis aligned with the major and minor axes of the ellipse, respectively and the $z$-axis is along the angular momentum vector $\mathcal{L}$, which is perpendicular to the orbital plane, the orbital motion of the secondary SMBH is given by
\begin{equation}
    x=R \cos f,  y= R \sin f,  z=0
\end{equation}
Next, we define a new coordinate system ($X,Y,Z$), in which the $Z$-axis is the observer's line of sight (L.O.S). In this system, the ($X,Y$) plane is called the plane of the sky. Below we define some geometry elements, which we also illustrate in \autoref{Fig:eccentric_orbit}: 
\begin{itemize}
    \item \emph{Inclination}: $i$ is the angle between the orbital plane and the plane of the sky, or equivalently the angle between the L.O.S. and the angular momentum vector, $\cos i = \hat{\mathcal{L}}\cdot\hat{Z}$. The binary orbit is edge-on when $i=90^{\circ}$ and face-on when $i=0^{\circ}$.
    \item \emph{Line of nodes}: The line along which the reference and the orbital plane intersect, with ascending/descending node the point of the orbit where the secondary crosses the reference plane moving upwards/downwards with respect to the observer's L.O.S.
    \item \emph{Longitude of the ascending node}: $\Omega_0$ is the angle between the $X$-axis and the ascending node. 
    \item \emph{Argument of periapse}: $w$ is the angle between the ascending node and the periapse.
\end{itemize}
Based on the above geometry, we can write the equations of motion of the secondary in the ($X,Y,Z$) system as follows
\begin{align}
    X&= R \left( \cos \Omega_0 \cos(w+f) - \sin \Omega \sin(w+f) \cos i \right)\nonumber\\
    Y&= R \left( \sin \Omega_0 \cos(w+f) + \cos \Omega \sin(w+f) \cos i \right)\nonumber\\
    Z&=R \sin(w+f) \sin i
\end{align}

Finally, we generalize the above for a higher mass ratio binary. In this case, both SMBHs orbit around the center of mass, which is either stationary (or moving with a constant velocity in a straight line). Here we will assume that the center of mass is not moving. Both orbits are eccentric with the same eccentricity, and the same period, but with different semi-major axes. The positions of the two SMBHs $R_1$ and $R_2$ at the opposite sides of the center of mass, with $R_1=M_2/(M_1+M_2)R$ and $R_2=M_1/(M_1+M_2)R$, where $R=R_1+R_2$ the distance between the two SMBHs. Additionally, the orbital velocities are $v_1=\dot{R_1}=q/(1+q)\dot{R}$ and $v_2=\dot{R_2}=1/(1+q)\dot{R}$.

For the electromagnetic signatures discussed below, one important quantity is the velocity along the line of sight $v_{\parallel}$. Given that the line of sight corresponds to the $Z$-axis, $v_{||}\propto\dot{Z}$, with $ v_{1||}=q/(1+q)\dot{Z}$, and $ v_{2||}=1/(1+q)\dot{Z}$, where
\begin{align}
\label{eq:vr_eccentric}
    \dot{Z}&=\dot{R}\sin(w+f)\sin i+R \dot{f}\cos(w+f)\sin i \nonumber\\
        &=\frac{2\pi}{P}\frac{a}{\sqrt{1-e^2}} e\sin f \sin(w+f)\sin i\nonumber\\
        &+\frac{2\pi}{P}\frac{a}{\sqrt{1-e^2}}(1+e\cos f)\cos(w+f)\sin i\nonumber\\
        &=\frac{2\pi a \sin i}{P\sqrt{1-e^2}}\left[\cos(w+f)+e (\sin f \sin(w+f)+\cos f \cos(w+f)\right]\nonumber\\   
        &=\left[\frac{G M_{\rm tot}}{a(1-e^2)}\right]^{1/2} \sin i\left[\cos(w+f)+e \cos w\right]
\end{align}
where we have used Kepler's law (\autoref{eq:KeplerLaw}), and the expressions for $\dot{R}$ and $R\dot{f}$ from \citet{2010exop.book...15M}.

\section{GW signals}
\label{sec:GWs}

GWs are space-time metric perturbations produced by the bulk motions of massive compact objects. To leading order, GWs are generated by an accelerated second-moment of a mass distribution, e.g., a binary system. The amplitude and phase of GW signals encode the time-variable properties of the emitting system. Here we consider the (directionally-dependent) timing deviations induced by GWs in the train of radio pulses from Galactic milli-second pulsars. Next, we examine the most generic waveform of an eccentric binary and link the GW signal to the orbital dynamics of the SMBHB. Finally, we briefly summarize how this method has been applied to PTA data returning astrophysically meaningful upper limits.

\subsection{PTA response to GWs}
PTAs rely on milli-second pulsars, which are very stably rotating neutron stars \citep{1968Natur.217..709H}. Their strong magnetic fields result in highly collimated radiation along the pulsar's magnetic field axis, which is typically misaligned with the rotation axis. As the pulsar rotates, a pulse is recorded by terrestrial radio telescopes every time a radiation beam sweeps across our line of sight. This gives rise to a periodic train of very precisely repeating radio pulses. High-precision timing experiments that use pulsars to search for GWs focus exclusively on pulsars with milli-second rotational periods, since they offer greater timing stability and fewer noise artifacts. 

A passing GW will interact with the Earth-pulsar baseline inducing deviations to the arrival time of radio pulses.
Specifically, we consider a pulsar with unit position vector $\hat{p}$, and a GW propagating in direction $\hat\Omega$. We note that $\hat\Omega$ is identical to the unit vector $\hat{Z}$ in the L.O.S. introduced above, but for consistency with the notation in the GW literature, we will keep $\hat\Omega$. The GW will induce fractional shifts to the arrival rate, $\nu$, of radio pulses with respect to the rate $\nu_0$, in the absence of GWs, according to
\begin{align} \label{eq:pulse_shift}
\frac{\nu_0-\nu(t)}{\nu_0} &= \frac{1}{2}\frac{\hat{p}^a \hat{p}^b}{(1+\hat\Omega\cdot\hat{p})}\left[h_{ab}(t,\hat\Omega) - h_{ab}(t_p,\hat\Omega) \right] \nonumber\\ 
    &= \frac{1}{2}\frac{\hat{p}^a \hat{p}^b}{(1+\hat\Omega\cdot\hat{p})}\Delta h_{ab},
\end{align}
where $a,b$ denote spatial components of vectors and tensors \citep{Detweiler1979}. Formally this calculation is an integral of the GW influence over the entire photon path, however in the short GW-wavelength limit of PTAs (wavelengths are much shorter than the distances to pulsars)\footnote{For a GW with $f_{\rm GW}=10$\,nHz, the GW wavelength is $\lambda_{\rm GW}=10^{16}\,\mathrm{m}<1\,\mathrm{pc}$, whereas the typical distances to pulsars are of order kpc.} we only retain effects from the limits of integration. This means that the influence of a GW on the pulse arrival rates is encoded by two stages of the emitting system's dynamics: $(1)$ at the time when the GW passes Earth $h_{ab}(t,\hat\Omega)$ (the \textit{Earth term}), and $(2)$ at the time when the GW passed the pulsar $h_{ab}(t_p,\hat\Omega)$ (the \textit{pulsar term}). The latter is always earlier, corresponding to a time $t_p \equiv t-L_p(1+\hat\Omega\cdot\hat{p})/c$, where $L_p$ is the distance to the pulsar.    

\begin{figure}
 \includegraphics[width=\columnwidth]{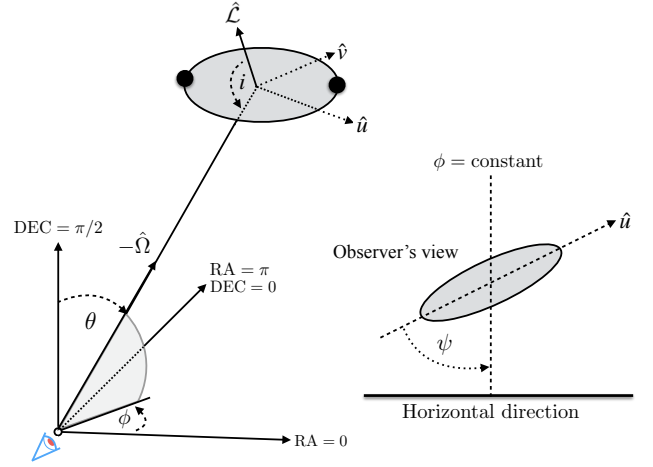}
 \caption{The orientation of a binary orbit as viewed from a coordinate system with the observer at the origin. The sky position of the binary is denoted by spherical polar $(\theta,\phi)$ angles, while the GW polarization angle is defined as the angle subtended between the GW basis vector $\hat{u}$ and a line of constant $\phi$, with $\hat{v}$ being another GW basis vector (see text for further details). The binary angular momentum vector is $\hat{\mathcal{L}}$, where the angle between this and the binary position vector defines the orbital inclination, $i$. Adapted from \citet{2016ApJ...817...70T}.}
 \label{Fig:gw_detector_angles}
\end{figure}

The metric perturbation for a GW propagating in direction $\hat\Omega$ at an arbitrary time can be written as the superposition of the two distinct GW-polarizations permitted in general relativity: 
the \textit{plus} and \textit{cross} modes,
\begin{equation}
    h_{ab}(t,\hat\Omega) = h_+(t) e^+_{ab}(\hat\Omega) + h_\times(t) e^\times_{ab}(\hat\Omega),
\end{equation}
where $e^{+,\times}_{ab}$ are basis tensors for these modes, offset by $\pi/4$ from each other, and named for how they deform a ring of test masses in the plane perpendicular to the direction of GW propagation. These basis tensors can be defined in terms a right-handed coordinate system that has $\hat{n}\equiv-\hat{\Omega}$ as one of the basis vectors. The other two basis vectors $\hat{u},\hat{v}$ are on the reference plane with $\hat{u}$ the unit vector towards the descending node and $\hat{v}$ orthogonal to it. More specifically%
\begin{align}
    \hat{n} \equiv &-\hat\Omega = \left(\sin\theta\cos\phi, \sin\theta\sin\phi, \cos\theta\right),\nonumber\\
    \hat{u} = &\,(\hat{n}\times\mathcal{\hat{L}}) / |\hat{n}\times\mathcal{\hat{L}}| \nonumber\\
    =&\left(\cos\psi\cos\theta\cos\phi - \sin\psi\sin\phi, \right.\nonumber\\
    &\left.\cos\psi\cos\theta\sin\phi + \sin\psi\cos\phi,\right.\nonumber\\ &\left.-\cos\psi\sin\theta\right),\nonumber\\
    \hat{v} = &\,\hat{u}\times\hat{n} \nonumber\\
    =&\left(\sin\psi\cos\theta\cos\phi + \cos\psi\sin\phi,\right. \nonumber\\
    &\left.\sin\psi\cos\theta\sin\phi - \cos\psi\cos\phi,\right.\nonumber\\ &\left.-\sin\psi\sin\theta\right),
\end{align}
where $(\theta,\phi) = (\pi/2 - {\rm DEC}, {\rm RA})$ denotes the sky-location of the SMBHB in spherical polar coordinates, and $\psi$ is the GW polarization angle that corresponds to the angle between $\hat{u}$ and the line of constant azimuth when the orbit is viewed from our coordinate system origin. In \autoref{Fig:gw_detector_angles} we show a representation of these angles from the point of view of an observer. With these basis vectors, the GW polarization basis tensors are
\begin{align} \label{eq:polbasis}
    e^+_{ab} = \hat{u}_a\hat{u}_b - \hat{v}_a\hat{v}_b, \nonumber\\
    e^\times_{ab} = \hat{u}_a\hat{v}_b + \hat{v}_a\hat{u}_b.
\end{align}

While \autoref{eq:pulse_shift} encodes the impact of a GW on the arrival rate of pulses, our observables are the times of arrival of these pulses. Hence it is more convenient to express the GW influence as an induced timing delay (or advance). The timing deviation induced by an individual GW signal is the accumulated fractional shift to the pulse arrival rate from the starting epoch of observations:
\begin{align}
    R(t) &\equiv \int_0^t dt'\,\left[\frac{\nu_0-\nu(t')}{\nu_0}\right] \nonumber\\ &=\frac{1}{2}\frac{\hat{p}^a \hat{p}^b}{(1+\hat\Omega\cdot\hat{p})} \int_0^t dt' \left[h_+(t') e^+_{ab}(\hat\Omega) + h_\times(t') e^\times_{ab}(\hat\Omega)\right.\nonumber\\
    &\left.\qquad\qquad\qquad\qquad\qquad - h_+(t'_p) e^+_{ab}(\hat\Omega) - h_\times(t'_p) e^\times_{ab}(\hat\Omega) \right] \nonumber\\
    &=\frac{1}{2}\frac{\hat{p}^a \hat{p}^b}{(1+\hat\Omega\cdot\hat{p})}e^+_{ab}(\hat\Omega)\int_0^t dt'\left[h_+(t')-h_+(t'_p)\right] \nonumber\\
    &\quad+\frac{1}{2}\frac{\hat{p}^a \hat{p}^b}{(1+\hat\Omega\cdot\hat{p})}e^\times_{ab}(\hat\Omega)\int_0^t dt'\left[h_\times(t')-h_\times(t'_p)\right] \nonumber\\
    &=F^+(\hat\Omega)\Delta s_+(t) + F^\times(\hat\Omega)\Delta s_\times(t),
\end{align}
where $s_{+,\times}(t) \equiv \int_0^t dt' h_{+,\times}(t')$ are the time-integrated polarization amplitudes, and $\Delta s_{+,\times}$ are the corresponding differences in Earth-term and pulsar-term delays. The functions $F^{+,\times}(\hat\Omega)$ encode the response of a given pulsar L.O.S. to the respective polarization modes of a GW propagating in direction $\hat\Omega$. These are defined by
\begin{equation}
\label{eq:antenna_response}
    F^{+,\times}(\hat\Omega) = \frac{1}{2}\frac{p^a p^b}{(1+\hat\Omega\cdot\hat{p})}e^{+,\times}_{ab}(\hat\Omega).
\end{equation}
The GW polarization amplitudes $h_{+,\times}$ are source-dependent. In the next section we present the waveforms for the most generic case of an eccentric binary system.

\subsection{Binary waveforms}
\label{sec:waveforms}

For the polarization amplitudes $h_{+,\times}$, we use the eccentric binary waveform from \citet{2016ApJ...817...70T}, which employs leading-order waveforms from  \citet{pm63} that are described in \citet{2004PhRvD..69h2005B}. Here we ignore higher order post-Newtonian terms, although we note that such waveforms exist, and are being actively implemented for PTA analysis \citep{2020PhRvD.101d3022S}. We also ignore BH spin, since the effects of the spin on the GW signal are unlikely to be observable by PTAs within the next decade \citep{2010PhRvD..81j4008S,2012PhRvL.109h1104M}. 

The time-dependent components of $s_{+,\times}$ can be computed analytically under the assumption that the system is not evolving over the observational baseline of the pulsar (on average $\sim10-20$ years) at the time when the GW passes the Earth (for the Earth term) and at the time at which the GW previously passed the relevant pulsar (for the pulsar term). This is a reasonable assumption given that PTAs are most sensitive to the early stages of the GW-driven evolution, thousands of years before the merger, when the binary evolution is extremely slow \citep{2010PhRvD..81j4008S}. Since we ignore the temporal evolution of binary parameters within the baseline of the Earth/pulsar terms, this means that time only appears in the definition of the binary's mean anomaly, where it is linear ($l(t) \approx l_0 + \omega t$, see \autoref{eq:mean_anomaly}). We modify the description in \citet{2016ApJ...817...70T} to allow for evolution of the pericenter direction over the much longer time difference between the pulsar- and Earth-term, which we parametrize with angle $\gamma$, described in further detail below \citep{1992MNRAS.254..146J}:
\begin{equation}\label{eq:pericenter_evolve}
    \dot\gamma\equiv \frac{d\gamma}{dt} = \frac{3\omega}{c^2} \frac{(G\omega M_{\rm tot})^{2/3}}{(1-e^2)} \left[ 1 + \frac{(G\omega M_{\rm tot})^{2/3}}{4c^2(1-e^2)}(26-15e^2) \right].
\end{equation}
Thus $\gamma(t) \approx \gamma_0 + \dot\gamma t$. 
At a given time, either during the baseline of the Earth- or pulsar-term, the time dependence of the GW-induced pulsar-timing delay caused by a single SMBHB is given by
\begin{align} \label{eq:splusscross-res}
    s_+(t) =& \sum_n -(1+\cos^2 i)[a_n S_{a,n}(t)-b_n S_{b,n}(t)] + (1-\cos^2 i)c_n, \nonumber\\
    s_{\times}(t) =& \sum_n 2\cos{i}[a_n C_{a,n}(t) - b_n C_{b,n}(t)],
\end{align}
where $n$ is an integer counting over harmonics of the binary's Keplerian orbital frequency. Even though the sum is formally over $[0,\ldots,\infty]$, in practice, accurate solutions can be computed by truncating the sum such that $n_\mathrm{max} \geq \ceil{3/(1-e)^{3/2}}$ \citep{2016ApJ...817...70T}. The harmonic-dependent functions and coefficients are given by
\begin{align}
    S_{a,n}(t) &= \frac{1}{2}\left[\frac{\sin[nl(t)-2\gamma(t)]}{n\omega-2\dot\gamma} + \frac{\sin[nl(t)+2\gamma(t)]}{n\omega+2\dot\gamma}\right], \nonumber\\
    S_{b,n}(t) &= \frac{1}{2}\left[\frac{\sin[nl(t)-2\gamma(t)]}{n\omega-2\dot\gamma} - \frac{\sin[nl(t)+2\gamma(t)]}{n\omega+2\dot\gamma}\right], \nonumber\\
    C_{a,n}(t) &= \frac{1}{2}\left[\frac{\cos[nl(t)-2\gamma(t)]}{n\omega-2\dot\gamma} - \frac{\cos[nl(t)+2\gamma(t)]}{n\omega+2\dot\gamma}\right], \nonumber\\
    C_{b,n}(t) &= \frac{1}{2}\left[\frac{\cos[nl(t)-2\gamma(t)]}{n\omega-2\dot\gamma} + \frac{\cos[nl(t)+2\gamma(t)]}{n\omega+2\dot\gamma}\right],
\end{align}
and
\begin{align} \label{eq:splusscross-coeffs}
    a_n=&\; -n\zeta\omega^{2/3}\left[J_{n-2}(ne)-2eJ_{n-1}(ne)+(2/n)J_n(ne)\right. \nonumber\\
    &\left.\vphantom{J_{n-2}(ne)-2eJ_{n-1}(ne)+(2/n)J_n(ne)}+2eJ_{n+1}(ne)-J_{n+2}(ne)\right]\nonumber\\
    b_n =& -n\zeta\omega^{2/3}\sqrt{1-e^2}\!\left[J_{n-2}(ne)-2J_n(ne)+J_{n+2}(ne)\right] \nonumber\\
    c_n =& \,\,\, (2/n)\zeta\omega^{-1/3}J_n(ne)\sin[nl(t)].
\end{align}

The parameters in the equations above have the following definitions: 
\begin{itemize}[leftmargin=15pt,labelwidth=15pt]
    \setlength\itemsep{0pt}
    \item \textit{Amplitude}: $\zeta=(G\mathcal{M}_z)^{5/3}/c^4 D_L$, where $D_L$ is the luminosity distance of the binary, $\mathcal{M}_z = (1+z)\mathcal{M}$ is the \textit{redshifted} chirp mass, and chirp mass is defined $\mathcal{M} = (M_1 M_2)^{3/5} / (M_1 + M_2)^{1/5}$.
    \item \textit{Pericenter angle:} $\gamma$ is an azimuthal angle measuring the direction of the system pericenter with respect to the vector $(\hat\Omega+\hat{\mathcal{L}}\cos i)/\sqrt{1-\cos^2 i}$, 
    which is orthogonal to the direction of the orbit's ascending node $-\hat{u}=\hat\Omega\times\hat{L}$. Given that the periapsis angle $w$ is referenced to the ascending node, the relationship between these two pericenter angles is $\gamma = w-\pi/2$. 
    \item The functions $J_{n}$ are n$^{\rm th}$-order Bessel functions.
\end{itemize}

If we ignore pericenter advance, and consider only circular binaries, these equations simplify significantly. When $e=0$ we have Bessel terms $J_0(0)=1$ and $J_{n>0}(0)=0$, such that the only remaining GW emission occurs at $f = 2f_{K,r}/(1+z)$. The term $c_n = 0$ for all harmonics, and only $a_2$ and $b_2$ are non-zero, taking values of $a_2 = b_2 = -2\zeta\omega^{2/3}$. Therefore the $s_{+,\times}$ terms reduce to
\begin{align} 
    s_+(t) &= \zeta\omega^{-1/3}(1+\cos^2 i)\sin(2\omega t + 2\Phi_0), \nonumber\\
    s_{\times}(t) &= 2\zeta\omega^{-1/3}\cos i\cos(2\omega t + 2\Phi_0),
\end{align}
where in this circular case the angles $l_0$ and $\gamma_0$ are often packaged together as $\Phi_0\equiv l_0+\gamma_0$ such that the signal model can be written more succinctly. 

Even though we assume that the binary is non-evolving over a pulsar observation baseline, it is necessary to model binary evolution during the (much longer) time lag between the pulsar-term and Earth-term. The Earth-term signal observed in each pulsar is the same up to the antenna response factors (\autoref{eq:antenna_response}), corresponding to a common orbital epoch of the SMBHB. But pulsar-terms are snapshots of the orbital dynamics that have a pulsar-specific lag behind the Earth-term by $t_e-t_p = L_p(1+\hat\Omega\cdot\hat{p})/c$. For distances of $\sim$~kiloparsecs the lag time can be thousands of years. This means that the pulsar-terms encode the past history of the binary, which can all be stitched together into a record of orbital evolution over $\sim$~thousands of years, despite the relatively short observation time of each pulsar. 

The orbital parameters of the source at the time of the pulsar-terms can be calculated by evolving the relevant Earth-term parameters backward in time using the evolution equations from  \citet{pm63} and \citet{,1964PhRv..136.1224P}
\begin{align} 
    \frac{d\omega}{dt} &= \frac{96c^6}{5G^2\mathcal{M}^2}\left(\frac{G\omega\mathcal{M}}{c^3}\right)^{11/3}\frac{1+\frac{73}{24}e^2+\frac{37}{96}e^4}{(1-e^2)^{7/2}},\nonumber\\
    \frac{de}{dt} &= -\frac{304c^3}{15G\mathcal{M}}\left(\frac{G\omega\mathcal{M}}{c^3}\right)^{8/3}e\frac{1+\frac{121}{304} e^2}{(1-e^2)^{5/2}},
\end{align}
and \autoref{eq:pericenter_evolve} for pericenter evolution. From the above equations, we see that GW emission causes a binary's orbital frequency to increase and its eccentricity to decrease, i.e. the emission of GWs tends to circularize the binary orbit \citep{1964PhRv..136.1224P}.

In PTA analysis, the inclusion of pulsar-terms in the SMBHB signal model is important, as it improves GW sky localization \citep{2010arXiv1008.1782C,2011MNRAS.414.3251L} and allows us to break the degeneracy between the chirp mass and distance in the amplitude variable $\zeta$. However, in practice, modeling the pulsar-terms presents challenges. 
The GW likelihood can be highly oscillatory when searching over the pulsar distances \citep{2010arXiv1008.1782C,2011MNRAS.414.3251L,2013CQGra..30v4004E}, which are currently poorly constrained. For instance, constraints on the distances for the vast majority of pulsars are much broader than a gravitational wavelength \citep{2012ApJ...755...39V,2018arXiv181206262M}.

\subsection{Gravitational wave constraints}

The above methodology has been applied in search campaigns for individually resolvable SMBHB signals by NANOGrav, EPTA, and PPTA collaborations with timing baselines that exceed a decade. 
All published PTA searches to date have focused on circular binaries, and have not considered the interplay/confusion with the aggregate signal from the GW background.  
However, both of these assumptions are being relaxed in ongoing NANOGrav and IPTA analyses. Below we summarize the most recent findings of these searches.

NANOGrav's most recent analysis used an $11$-year dataset which includes $34$ pulsars \citep{2018ApJS..235...37A}. For an all-sky search that averages over the array's sensitivity to different sky locations, the strain amplitude of a putative SMBHB was limited to 
$h_0 \leq 7.3\times 10^{-15}$
at the array's most sensitive GW frequency of $f = 8$~nHz \citep{NANOGrav_CWs_2018}. In the most sensitive sky location, they were able to rule out binaries with chirp mass $\mathcal{M} > 10^9 (10^{10}) M_\odot$ within $120$~Mpc ($5.5$~Gpc) at $95\%$ credibility. The EPTA Data Release 1 contained timing information of $41$ pulsars whose timespans range from $7-24$~years \citep{Desvignes2016} and provided an upper limit of $h_0 \leq 10^{-14}$ for frequencies $5~\mathrm{nHz} < f < 7~\mathrm{nHz}$ \citep{Babak2016}. EPTA limits exclude the presence of binaries with chirp mass $\mathcal{M} > 10^9 (10^{10}) M_\odot$ out to a distance of about $25$~Mpc ($1$~Gpc). Finally, the PPTA Data Release 1 contained timing information of $20$ pulsars with $\sim6$~year baseline \citep{2013PASA...30...17M}, which delivered constraints of $h_0 \leq 1.7\times 10^{-14}$ at $f=10$~nHz that exclude the presence of circular binaries with $\mathcal{M}=10^9 M_\odot$ any closer than $100$~Mpc \citep{Zhu2014}. These limits vary across the sky by a factor of a few to an order of magnitude (depending on the specific PTA) due to the inhomogeneous spatial distribution of monitored pulsars, which leads to an anisotropic signal response.

Beyond these blind, all-sky, source-agnostic searches, it is also possible to employ restrictive parameter priors when targeting specific prominent galaxy clusters or binary candidates identified with electromagnetic data. For instance, NANOGrav's $11$-year dataset was used to show that there are no SMBHBs with $\mathcal{M}\geq 1.6\times 10^9 M_\odot$ emitting GWs with frequencies $2.8-317.8$~nHz in the Virgo cluster. Likewise, NANOGrav performed a recent targeted analysis of the galaxy $3$C$66$B, which was proposed to host a SMBHB because of a $1.05$~year periodicity inferred from astrometric observations  \citep{2003Sci...300.1263S}, thereby placing its potential GW emission in the PTA band.
The first analysis of this candidate \citep{Jenet2004} was one of the first ever successes of multi-messenger astronomy, where data from the pulsar B$1855$$+$$09$ constrained the chirp mass of the putative binary to be $\mathcal{M}\leq 7\times 10^9 M_\odot$. NANOGrav's recent analysis using its $11$-year dataset found that if the period is constrained to within an order of magnitude, and subsequently included as a prior in the GW analysis, then this leads to significantly improved constraints on the chirp mass $\mathcal{M}\leq 1.65\times 10^9 M_\odot$ \citep{2020ApJ...900..102A}.

Significant constraints were produced not only for binary candidates but also for the general population of nearby massive galaxies. NANOGrav recently compiled a target list of $216$ massive galaxies (from a comprehensive catalog of $\sim$45,000 galaxies within 500\,Mpc) that fall within its sensitivity volume, placing multi-messenger constraints on the chirp mass and mass ratio of potential binaries in each of those galaxies. In several of these galaxies only unequal binaries are allowed given NANOGrav's upper limits,
and $19$ systems could be constrained with the same precision as a tentative SMBHB in our own Milky Way \citep{2021arXiv210102716A}.

\section{EM signals}
\label{sec:em_signals}
Binaries likely reach the GW regime surrounded by an abundant gas reservoir, since galaxy mergers are expected to funnel significant amounts of gas to the central regions of the post-merger galaxies \citep{BarnesHernquist1992,2005ApJ...620L..79S,2006ApJ...645..986R}. As a result, binaries may produce bright electromagnetic signatures \citep{2002ApJ...567L...9A,2009ApJ...700.1952H,2012MNRAS.420..705T,2021arXiv210903262B}. In this section, we describe how these signatures are related to the orbital elements of the binary (so that they can be directly linked to the GW signal for joint multi-messenger observations). Since most of the phenomenology was derived by hydrodynamical simulations, we start by presenting the status of theoretical studies. We also present a brief summary of the observational efforts motivated by these signatures which led to the detection of a few hundreds of candidates.

\subsection{Circumbinary disks}
\label{sec:disks}
Gas in the vicinity of the binary is expected to settle in a rotationally-supported circumbinary disk \citep{Barnes2002}. The disk exchanges energy and angular momentum with the binary, which means that the disk, in addition to fuelling the binary's electromagnetic emission, plays a crucial role in the orbital evolution of the system. Initially, a gaseous disk was invoked as a solution to the final parsec problem \citep{2000ApJ...532L..29G}, since it was thought that it catalyzes the binary's orbital decay \citep{2002ApJ...567L...9A, 2008ApJ...672...83M,2009MNRAS.393.1423C,2012A&A...545A.127R}. However, recent simulations have brought this into question suggesting that interactions with the circumbinary disk can expand the binary orbit at least for certain binary and disk parameters \citep{2017MNRAS.466.1170M,2019ApJ...871...84M,2019ApJ...875...66M,2020ApJ...901...25D}. 

Even though the effects of gas on binary evolution are still unclear, simulations have converged on some generic features. First, the binary clears the central region of the disk, creating a cavity of low-density gas \citep{AL96,2008ApJ...672...83M,2013MNRAS.436.2997D,2014ApJ...783..134F}. However, this cavity is not completely devoid of gas, which would hinder electromagnetic emission. The dynamic interplay between the binary and the edge of the circumbinary cavity results in quasi-periodic mass transfer to the SMBHs \citep{2008ApJ...672...83M,2013MNRAS.436.2997D,2012A&A...545A.127R,2014ApJ...783..134F}.
As gaseous streams enter the cavity, some of the infalling material becomes bound to the SMBHs, and persistent mini-disks are formed \citep{2014ApJ...783..134F,Ryan2017}. Additionally, mass accretion is higher onto the secondary SMBH because it moves closer to the edge of the cavity.
The above features give rise to an array of electromagnetic signatures that may allow us to detect SMBHBs at sub-parsec separations, where the two SMBHs cannot be individually resolved through current means..

At larger sub-parsec separations, the gas that becomes bound to each SMBH may produce strong broad emission lines. In this case, the orbital motion of the binary may be detectable via time-dependent Doppler shifts of these broad emission lines in some quasars \citep{gaskell96}. These broad-emission-line shifts, measurable relative to the narrow emission lines that are emitted on galaxy-size scales, are periodically modulated at the orbital period of the binary. For this signature, either one or both of the SMBHs need to be active, and the broad-line regions (BLRs) associated with them need to be distinct, although potentially partially truncated \citep{runnoe15,DeRosa2019}. 
As the binary orbit shrinks, the broad line regions, which typically extend to a few light years \citep{2019ApJ...887...38G,2021ApJ...915..129K}, are truncated to the degree that they cannot produce broad emission lines \citep{kelley21}. 

In this regime, the most promising signature is periodic photometric variability. As mentioned above, the binary perturbs the circumbinary disk leading to periodic leakage of gas into the cavity \citep{2008ApJ...672...83M,2013MNRAS.436.2997D}, which may be translated to periodic brightness fluctuations of the binary. Periodic variations may be observed at the orbital period of the binary or at a few times the orbital period of the binary, depending on the binary's mass ratio \citep{2013MNRAS.436.2997D,2014ApJ...783..134F,2021arXiv210312100N}, and the eccentricity \citep{2017MNRAS.466.1170M} discussed below. The periodicity coincides with the orbital period for unequal-mass binaries, whereas for relatively equal-mass binaries, the dominant periodic component is expected at a several ($\sim5-8$) times the orbital period \citep{2013MNRAS.436.2997D,2014ApJ...783..134F,2017MNRAS.466.1170M}. This timescale corresponds to the orbital period of a hotspot (over-density) formed at the edge of the circumbinary disk \citep{Shi+12,2014ApJ...783..134F}. The exact mass-ratio where the transition occurs is still uncertain and depends on the simulation setup \citep{2013MNRAS.436.2997D,2014ApJ...783..134F, 2020ApJ...889..114M}. 

Beyond periodic accretion, relativistic Doppler boosting can also produce periodic photometric variability \citep{Dorazio2015Nature,Tang+2018}. For compact, milli-parsec binaries, the SMBHs (and thus the mini-disks around them) move at a fraction of the speed of light, and effects of special relativity are inevitable. 
For instance, the secondary mini-disk -- which is typically more luminous and the one moving faster -- 
will appear brighter when it is moving towards the observer and dimmer when it is moving away. Relativistic Doppler boosting will likely dominate the photometric variability for unequal mass-binaries and for orbits that are relatively close to edge-on ($i=90^{\circ}$). 
Doppler boosting can produce significant periodicity, even if the rest-frame luminosity of the mini-disks is constant, which is expected for very unequal-mass binaries ($q<0.05$). Low mass ratio binaries do not open a cavity, but instead open an annular gap at the orbit of the secondary. The Doppler boost variability (period, phase, amplitude) can be easily tied to the orbital geometry and dynamics of the binary. In addition to Doppler boost variability, in unequal mass binaries with orbits close to edge-on, periodic self-lensing flares may also be observed \citep{2018MNRAS.474.2975D,2020MNRAS.495.4061H}. 

The above conclusions were derived by simulations, in which the binary is moving in a circular orbit. However, the interaction of eccentric orbits with the circumbinary disk has received a lot of attention recently. The interaction of the binary with the circumbinary disk tends to circularize the binary for low eccentricity binaries ($e\leq0.1$), whereas it drives it to $e\sim 0.4$ for all other eccentricities \citep{2009MNRAS.393.1423C,2012JPhCS.363a2035R,2021ApJ...909L..13Z,2021ApJ...914L..21D}. The accretion rate is periodically modulated at the orbital period of the binary and shows repeating bursts \citep{2009MNRAS.393.1423C,2016ApJ...827...43M,2017MNRAS.466.1170M,2021ApJ...909L..13Z}.
Additionally, eccentricity was included in Doppler boost models giving rise to periodic but very non-sinusoidal profiles \citep{2020MNRAS.495.4061H}.

\subsection{Periodic Doppler shifts of the broad emission lines} \label{sec:radial_velocity_theory}

The main observables of the radial velocity method are velocity offsets of the broad emission lines of quasars relative to the narrow lines, and their time-dependent changes.
This approach is applicable for a limited range of separations, i.e. sub-pc separations corresponding to periods of decades to a few hundred years \citep{eracleous12}. At wide separations, the binary must be gravitationally bound \citep{kelley21} and the velocity offset and acceleration due to bulk orbital motion of the BLR must be perceptible \citep{eracleous12,pflueger18,kelley21}. At close separations, the binary must still be wide enough to produce relatively normal, luminous broad emission lines similar to regular (single-SMBH) quasars \citep{runnoe15,kelley21}.
One of the strengths of this technique is that it can place limits on the period, separation and mass ratio of the tentative binary with observations covering only a fraction of an orbital cycle \citep{Runnoe2017}.

\subsubsection{Circular orbits}

A treatment of the radial velocity method under the assumption of circular orbits has been presented in a number of previous studies \citep[e.g.,][]{bogdanovic09,eracleous12,shen13,runnoe15,pflueger18,kelley21}. The observed velocity offset due to orbital motion projected along the L.O.S. is $v_\parallel = v\,\sin i\sin(\omega t+\Phi_0)$, where $v$ is the orbital velocity of the primary or secondary expressed in \autoref{eq:velocities}. For instance, if we consider the motion of the secondary SMBH, the observed velocity offset and acceleration are given by
\begin{align}\label{eq:vobs}
    v_{2\parallel} &= v_2\,\sin i\sin(\omega t+\Phi_0) \nonumber \\
    &= \frac{1}{1+q}\left(\frac{2\pi GM_{\rm tot}}{P}\right)^{1/3}\sin i\sin(\omega t+\Phi_0),
\end{align}
\begin{align}\label{eq:vobs}
    \frac{d v_{2\parallel}}{d t} &= v_2\,\sin i\,\frac{d}{d t}\left\{\sin(\omega t+\Phi_0)\right\} \nonumber \\
    &= \frac{1}{1+q}(GM_{\rm tot})^{1/3} \left(\frac{2\pi}{P}\right)^{4/3}  \sin i\cos(\omega t+\Phi_0),
\end{align}
where we have substituted $\omega = 2\pi/P$ after the derivative. The observable velocity offset and acceleration of the primary are a factor of $q$ larger and out of phase by $\pi$ compared to the secondary.

\subsubsection{Eccentric orbits}
Searches for SMBHBs with the radial velocity method have focused on circular binary orbits because there are typically not enough data points in the radial velocity curve to constrain the many additional free parameters of eccentric orbits.
It is only recently that eccentric orbits have been introduced to model a binary candidate in the context of quasars with double-peaked emission lines \citep[e.g.,][]{doan20}. 
The receding SMBH produces a redshifted emission line in a quasar spectrum, while the approaching SMBH would produce a blueshifted emission line, with the L.O.S. velocity $v_{\parallel}$ related to \autoref{eq:vr_eccentric}. The relative velocity of the redshifted and blueshifted components is equal to the mass ratio (or the inverse of it) depending on whether the primary or secondary SMBH is the receding body.  Additionally, many tools have been developed for addressing this problem in the context of radial velocity searches for extrasolar planets \citep[e.g.,][]{ford05,fulton18}.

\subsection{Periodic Variability due to relativistic Doppler boost}\label{sec:doppler_boost}

For milli-parsec separated binaries in the GW-dominated regime, the orbital velocities of the SMBHs are a few percent of the speed of light. For instance, for a circular binary with mass $M_{\rm tot}=10^9M_{\odot}$, mass ratio $q=0.25$ and period $P=1$\,yr, the velocity of the secondary is $v_2\sim0.08 c$. Therefore, any luminosity associated with the fast moving SMBHs (e.g., emission from the mini-disks) will be affected by relativistic Doppler boosting. Below we derive step-by-step the mathematical formalism of this model.

Let's assume a source with specific intensity $I_{\nu}$ at photon frequency $\nu$, moving with relativistic velocity $v$. In special relativity the photon phase-space density $I_\nu/\nu^3$ is Lorentz invariant, since all observers will count the same number of photons. This means that
\begin{equation}
    \frac{I_\nu^{\rm obs}}{\left(\nu^{\rm obs}\right)^3}=\frac{I_\nu^{\rm em}}{\left(\nu^{\rm em}\right)^3}.
\end{equation}
The observed frequency of the photons will be modified due to relativistic motion as $\nu^{\rm obs} = \mathcal{D}\nu^{\rm em}$, where $\mathcal{D}$ is the Doppler factor that depends on $\beta=v/c$ as
\begin{equation}
    \mathcal{D}=\left[\frac{1-\beta_\parallel}{\sqrt{1-\beta^2}}\right]^{-1} \simeq\sqrt{\frac{1+\beta_{\parallel}}{1-\beta_{\parallel}}}, 
\end{equation}
where $\beta_{\parallel}=v_{\parallel}/c$ is the line-of-sight component of $\beta$.

Assuming that the emitted radiation has a
power-law spectrum with a spectral index $\alpha_{\nu}$, i.e., $I_{\nu}=K \nu^{\alpha_{\nu}}$, the observed specific intensity will be
\begin{equation}\label{eq:Inu}
    I_{\nu}^{\rm obs}=\left(\frac{\nu^{\rm obs}}{\nu^{\rm em}}\right)^3K\left(\nu^{\rm em}\right)^{\alpha_{\nu}}=
    \mathcal{D}^{3-\alpha_{\nu}}I_{\nu,0},
\end{equation}
where $I_{\nu,0}$ is the stationary (i.e. non boosted) specific intensity of the source at the observed photon frequency.
Additionally, the monochromatic flux $F_{\nu}$ of a uniformly emitting source 
is related to the specific intensity as $F_{\nu}=\pi I_{\nu}$, and thus we can express \autoref{eq:Inu} in terms of fluxes, i.e., $F_{\nu}^{\rm obs}=\mathcal{D}^{3-\alpha_{\nu}}F_{\nu,0}$. Taking the first order Taylor expansion in $\beta_{\parallel}$, we have 
\begin{align}
    \mathcal{D}^{3-\alpha_{\nu}} &= \left(\frac{1+\beta_{\parallel}}{1-\beta_{\parallel}}\right)^{(3-\alpha_{\nu})/2} \nonumber\\
    &\simeq \left(1+2\beta_{\parallel}\right)^{(3-\alpha_{\nu})/2} \nonumber\\
    &\simeq 1+(3-\alpha_{\nu})\beta_{\parallel}.
\end{align}
Substituting the above equation, the observed flux becomes
\begin{equation}\label{eq:DB_final}
    F_{\nu}^{\rm obs}=F_{\nu,0}+(3-\alpha_{\nu}) v_{\parallel} F_{\nu,0}/c = F_{\nu,0}\left(1+\Delta F\right),
\end{equation}
where $\Delta F = (3-\alpha_{\nu}) v_{\parallel}/c$.

\subsubsection{Circular orbits}
\label{sec:DB_simplest}
\begin{figure} 
    \begin{subfigure}{\columnwidth}
    \centering
     \includegraphics[width=\textwidth, trim={25cm 13cm 3 0},clip]{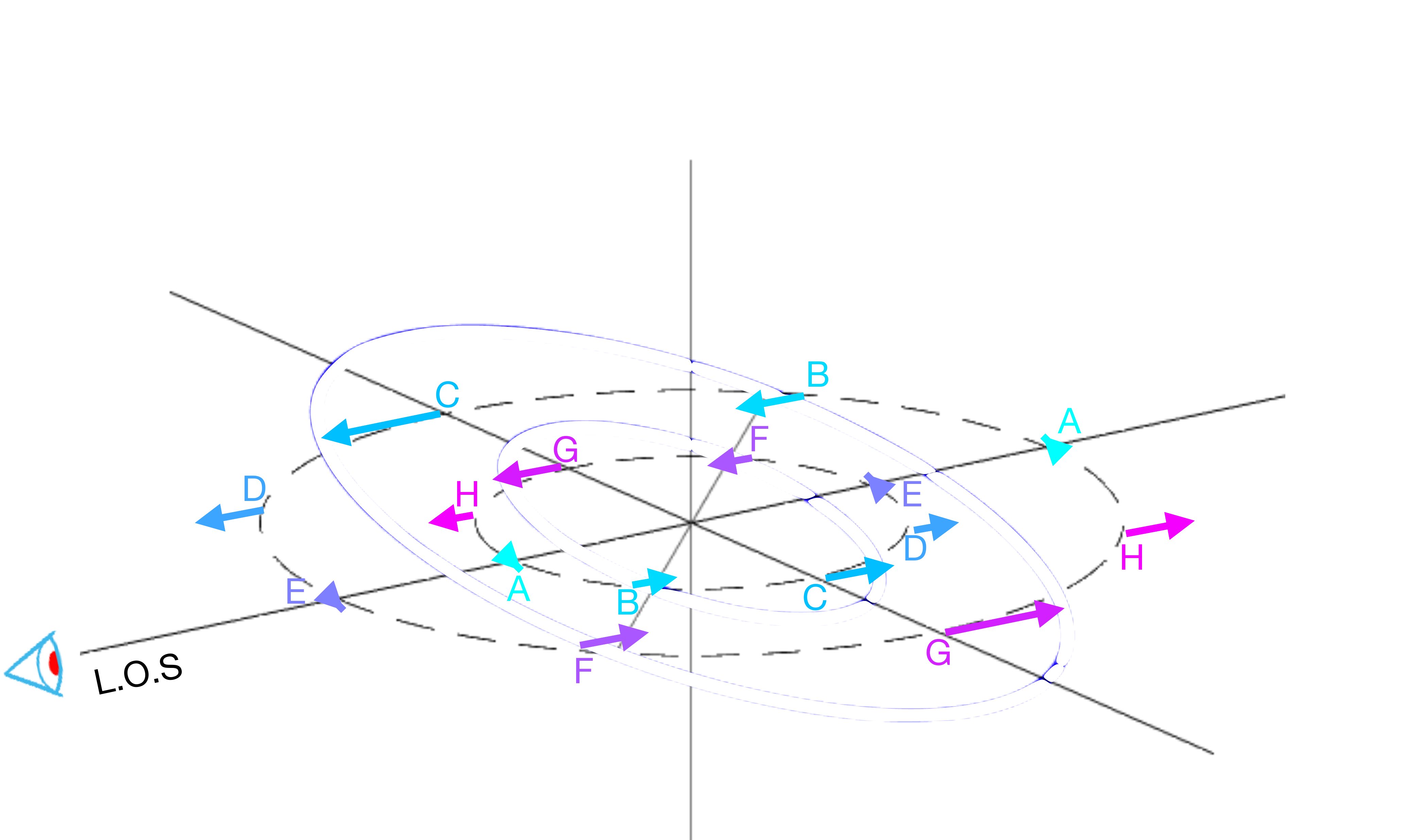}
    \end{subfigure}
    \begin{subfigure}{\columnwidth}
    \centering
     \includegraphics[width=\textwidth]{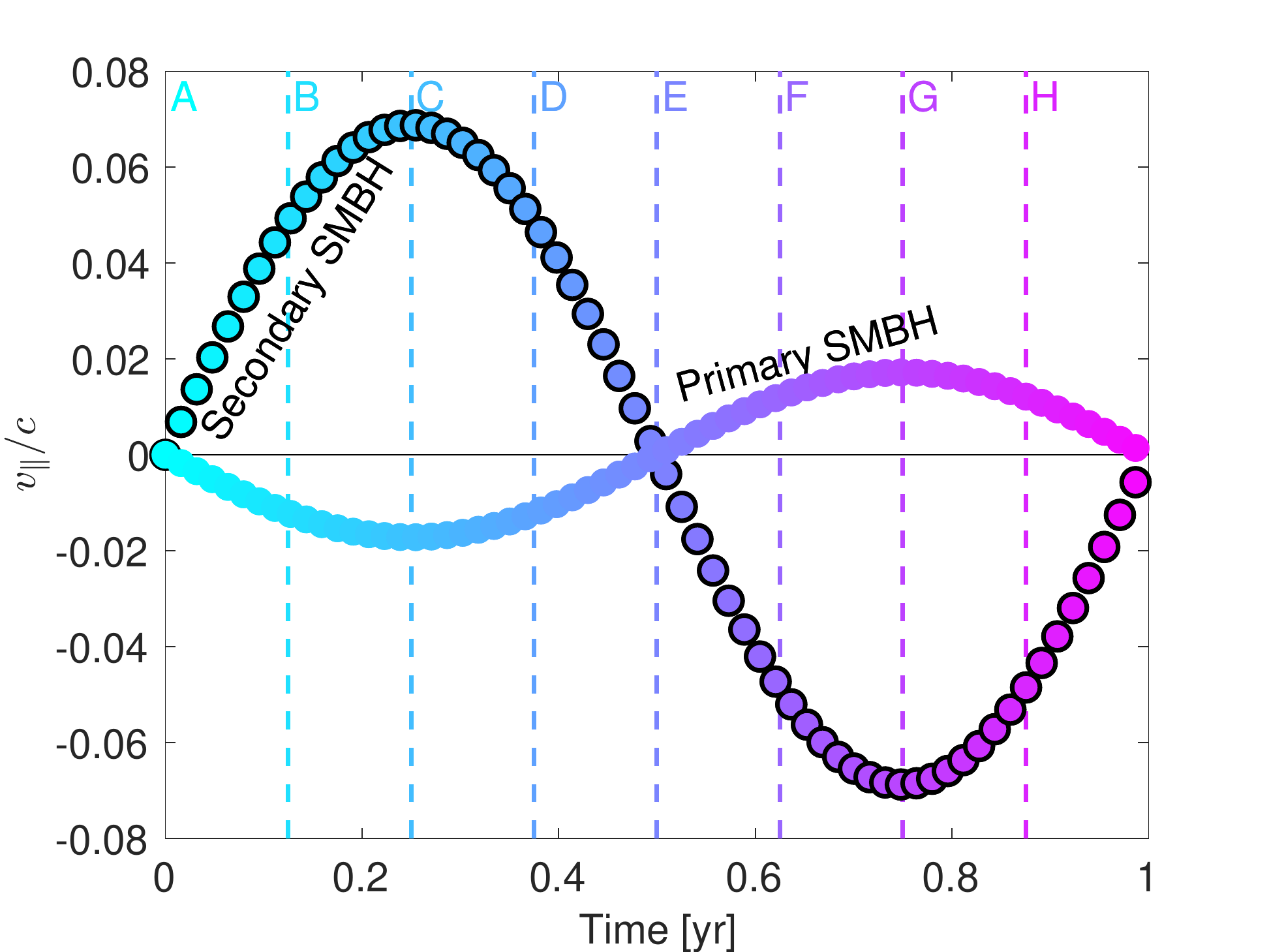}
    \end{subfigure}
\caption{Orbital velocity of the primary (open markers) and secondary (markers with black outline) SMBHs projected in the L.O.S. for the fiducial binary ($M_{\rm tot}=10^9M_{\odot}$, mass ratio $q=0.25$ and period $P=1$\,yr and $i=60^{\circ}$). The vertical lines indicate distinct times during the period of the binary, and correspond to the respective orbital configuration shown in the top panel.} 
\label{fig:V_LOS}
\end{figure}

In the simplest model of relativistic Doppler boosting that has been assumed in most previous papers \citep{Dorazio2015Nature,2018MNRAS.476.4617C,2019MNRAS.485.1579K,2020MNRAS.496.1683X}, there are several key assumptions:
\begin{enumerate}[leftmargin=*]
    \item[(1)] The (faster moving) secondary mini-disk is more luminous compared to the primary, and dominates the variability.
    \item[(2)] The intrinsic  mini-disk luminosity does not vary over time.
    \item[(3)] The binary is in a circular orbit.
    \item[(4)] The period of the binary does not evolve over the timespan of observations.
\end{enumerate}
For such a binary, the Doppler boost variability is given by
\begin{align}
\label{eq:DB_ciruclar}
    F_{\nu}&=F_{\nu,0}^{\rm sec}\left[1+(3-\alpha_{\nu}) v_{2 \parallel}/c\right]\nonumber\\
    &=F_{\nu,0}^{\rm sec}[1+(3-\alpha_{\nu})v_2\sin i\sin(\omega t+\Phi_0)/c],
\end{align}
where $F_{\nu,0}^{\rm sec}$ is the stationary luminosity of the secondary mini-disk.  
Since the variability strongly depends on the line-of-sight velocity, in \autoref{fig:V_LOS}, we show these velocities both for the primary and secondary SMBHs for the fiducial binary mentioned above ($M_{\rm tot}=10^9M_{\odot}$, mass ratio $q=0.25$ and period $P=1$\,yr and $i=60^{\circ}$).

The above model has been used repeatedly not only for its simplicity, but also because its underlying assumptions are reasonable. We have already mentioned in \S~\ref{sec:waveforms} that orbital evolution is slow for binaries in the PTA band. Additionally, the secondary SMBH moves faster (as expected from \autoref{eq:velocities}), and since the luminosity of the secondary mini-disk is higher it is reasonable to assume that the primary mini-disk may not contribute significantly. However, some of the above assumptions may not always hold. In \S~\ref{sec:discussion} we explore how relaxing the above assumptions impacts the Doppler boost variability, and any potential deviations from the sinusoidal variability delivered by the simplest scenario here.



\subsection{Periodic variability due to periodic accretion}\label{sec:periodic_accretion}

As the binary periodically perturbs the circumbinary disk, the accretion rate onto the binary is periodically modulated and this can be translated to periodic lightcurve variability. For unequal mass binaries ($0.05<q<0.3-0.5$), the dominant periodic component is at the orbital period of the binary. Since in this regime Doppler boost variability is expected to be significant, the variability produced from periodic accretion likely co-exists with relativistic Doppler boost \citep{Tang+2018}. The interplay between the two mechanisms has not been adequately explored in the literature and it is unclear whether one mechanism is more dominant than the other. If Doppler boosting is dominating the variability, as discussed above, the variability is expected to be smooth and quasi-sinusoidal, and the electromagnetic signal can be directly connected to the binary orbit and thus the GW signal (see \S~\ref{sec:mma}). The contribution of periodic accretion may lead to perturbations in the luminosity of the mini-disks, as the streams periodically hit the mini-disks -- we further explore this effect in \S~\ref{sec:DB_variable_lum}. On the other hand, if periodic accretion is the dominant component, the variability will likely be less sinusoidal, e.g., repeating bursts. In that case, the phase and the amplitude of the electromagnetic signal cannot be linked to the orbital dynamics of the binary, and thus the connection with the GW signal relies only on the common orbital period (see \S~\ref{sec:mma}).

For binaries close to equal mass, e.g., $q>0.3-0.5$ (although the precise mass ratio for the transition is not well constrained), the dominant periodic component is at several ($\sim5-8$) times the orbital period \citep{2013MNRAS.436.2997D,2014ApJ...783..134F}. This may be connected to brightness variations of an overdense hotspot of gas that builds up along the wall of the circumbinary disk. Simulations show the binary exhibiting saw-tooth variability on these timescales, with additional less prominent components at orbital and half-orbital periods \citep[e.g.,][]{2020ApJ...901...25D,Bowen_2017,2014ApJ...783..134F,2013MNRAS.436.2997D,2019ApJ...871...84M}.
A representative template for this accretion behavior is suggested by \citet{2020ApJ...901...25D}, which we augment here with orbital and half-orbital modulations in the fractional binary accretion rate:
\begin{align}
    \frac{\dot{M}(t)}{\dot{M}_0} = 1 -\,\, & a_\mathrm{lump}\tan^{-1}\left[\tan(\phi_\mathrm{lump}+ \omega t/5)\right] + \nonumber\\
    & a_\mathrm{b}\sin(\phi_\mathrm{b} + \omega t) +  a_\mathrm{hb}\sin(\phi_\mathrm{hb} + 2\omega t),
\end{align}
where the first variable term reproduces the sawtooth temporal behavior on timescales of the hotspot period ($\sim5P$) that is observed in some simulations. The other variability terms modulate the accretion rate (and thus the lightcurve behavior) on the binary's orbital period as well as half of its orbital period, respectively. Each term is given its own phase that may or may not be connected to the phase of the binary dynamics. As a result, linking the properties of this kind of electromagnetic signal to the GW signal via the binary orbital dynamics is less straightforward. The phase and amplitude of the hotspot are not directly connected to the orbital dynamics, and even the period harmonic connection is tentative. However, since the additional periodic components are at the orbital and half-orbital period, it is likely that these can facilitate the connection of the signals. This is especially true if the variability at the orbital period in fact arises from the Doppler boosting effect. 

We emphasize that an improved theoretical understanding of the systems through more detailed simulations is necessary. This is particularly important in the high mass-ratio regime ($q>0.3-0.5$), as equal mass binaries produce the strongest GW signals and thus can be detected by PTAs at larger distances.

\subsection{Constraints from electromagnetic observations}

The above signatures have been extensively used for systematic searches of binary candidates in spectroscopic and photometric data, revealing
several promising candidates, which we summarize below.

Initial searches for shifted broad emission lines focused on quasars with double-peaked broad emission lines, as the consequence of the scenario where both black holes are active \citep{gaskell88}. As we mentioned in \S~\ref{sec:radial_velocity_theory}, these are no longer considered strong candidates \citep[e.g.,][ and referneces therein]{eracleous97,doan20}. 
More recent searches have focused on the population of quasars with single-peaked broad emission lines with velocity shifts, identifying of order $10^2$ 
candidates. Many of these candidates were selected from large-area spectroscopic surveys searching for quasars with substantial (1000~km~s$^{-1}$) velocity offsets of their broad H$\beta$ lines \citep{tsalmantza11,eracleous12,liu14}. Subsequent long-term monitoring \citep{decarli13,runnoe15,Runnoe2017,guo19} and  theoretical modeling \citep{nguyen20} have provided radial velocity curves and constraints on the binary properties. Additionally, when the two SMBHBs are aligned with the observer, they do not have L.O.S. velocities and thus do not produce shifted lines. Such candidates have been found from measuring accelerations in their broad H$\beta$ and Mg~\textsc{ii}~$\lambda2798$ lines in multi-epoch spectroscopy \citep{shen13,ju13,Wang2017}.

The search for binaries at smaller separations based on AGN periodicity has seen rapid developments in the last five years.
The advent of large time-domain surveys has provided exceptional datasets for this type of searches and $\sim150$ candidates have been identified \citep{DeRosa2019}.
In particular, \citet{2015MNRAS.453.1562G} analyzed a sample $\sim$245,000 quasars from the Catalina Real-time Transient Survey (CRTS) and identified significant evidence for periodicity in 111 quasars. \citet{Charisi2016} performed a systematic search in a sample of $\sim$35,000 quasars from the Palomar Transient Factory (PTF) and detected 33 candidates. One additional candidate was found in a small sample of 9000 quasars from the Panoramic Survey Telescope and Rapid Response System (Pan-STARRS; \citealt{2019ApJ...884...36L}), while five more candidates were discovered in an even smaller sample of 625 AGN from the Dark Energy Survey (DES; \citealt{2020MNRAS.499.2245C}).

Identifying periodicity in quasars is challenging and the samples of candidates likely contain false detections \citep{2021arXiv211007465W}. For instance, the population of SMBHBs extrapolated from the population of periodic candidates \citep{2015MNRAS.453.1562G,Charisi2016} would produce a GW background inconsistent with upper limits from PTAs \citep{Sesana2018}. We note that all the above studies included the intrinsic quasar variability in the statistical analysis, modeling the quasar variability with the most successful model, namely the Damped Random Walk model. The contamination with false positives is expected due to a combination of the stochastic noise \citep{2016MNRAS.461.3145V,2021arXiv211007465W}, the relatively long periods compared to the available baselines, which do not allow us to observe many cycles and the relatively sparse data of the optical time-domain surveys. 

The uncertainty regarding the binary nature of the candidates has led to many searches for additional signatures. The Doppler boost model described above has a robust prediction, which can be tested with multi-wavelength data; the luminosity in two distinct bands arising in Doppler-boosted mini-disk (e.g., optical, UV, X-rays) should vary in tandem, but with amplitude which depends on the spectrum of the respective bands \citep{Dorazio2015Nature,2018MNRAS.476.4617C,2020MNRAS.496.1683X}. The broad emission lines, likely associated with the circumbinary disk, are also expected to respond periodically to the presence of a SMBHBs \citep{2021A&A...645A..15S,2021ApJ...910..101J}. Additionally, as mentioned above, if the periodicity is due to periodic accretion, the variability may contain multiple periodic components with a characteristic frequency pattern \citep{2015MNRAS.454L..21C,2015MNRAS.452.2540D}. The signature of the binary may also be imprinted on the IR emission from the dust in the AGN \citep{2015ApJ...814L..12J,2017MNRAS.470.1198D}, in the X-ray spectrum of the source \citep{2020ApJ...900..148S} and on the radio jets \citep{Kun2015,Mohan2016}. A subset of the candidates show one or more of the above promising signatures. However, the evidence remains inconclusive, and further monitoring is required to confirm (or disprove) their binary nature

\section{Multi-messenger Observations}
\label{sec:mma}
In the previous two sections we have presented how the GW and electromagnetic signatures of SMBHBs are connected with the orbital dynamics and geometry of the binary. From the above it appears that the Doppler boost signature (both for circular and eccentric binaries) is the most promising signature for multi-messenger observations, because the electromagnetic emission (amplitude, phase, period) is directly linked to the orbital geometry of the binary and thus can be linked to the GW signal.

\begin{figure} 
    \centering
     \includegraphics[width=\columnwidth]{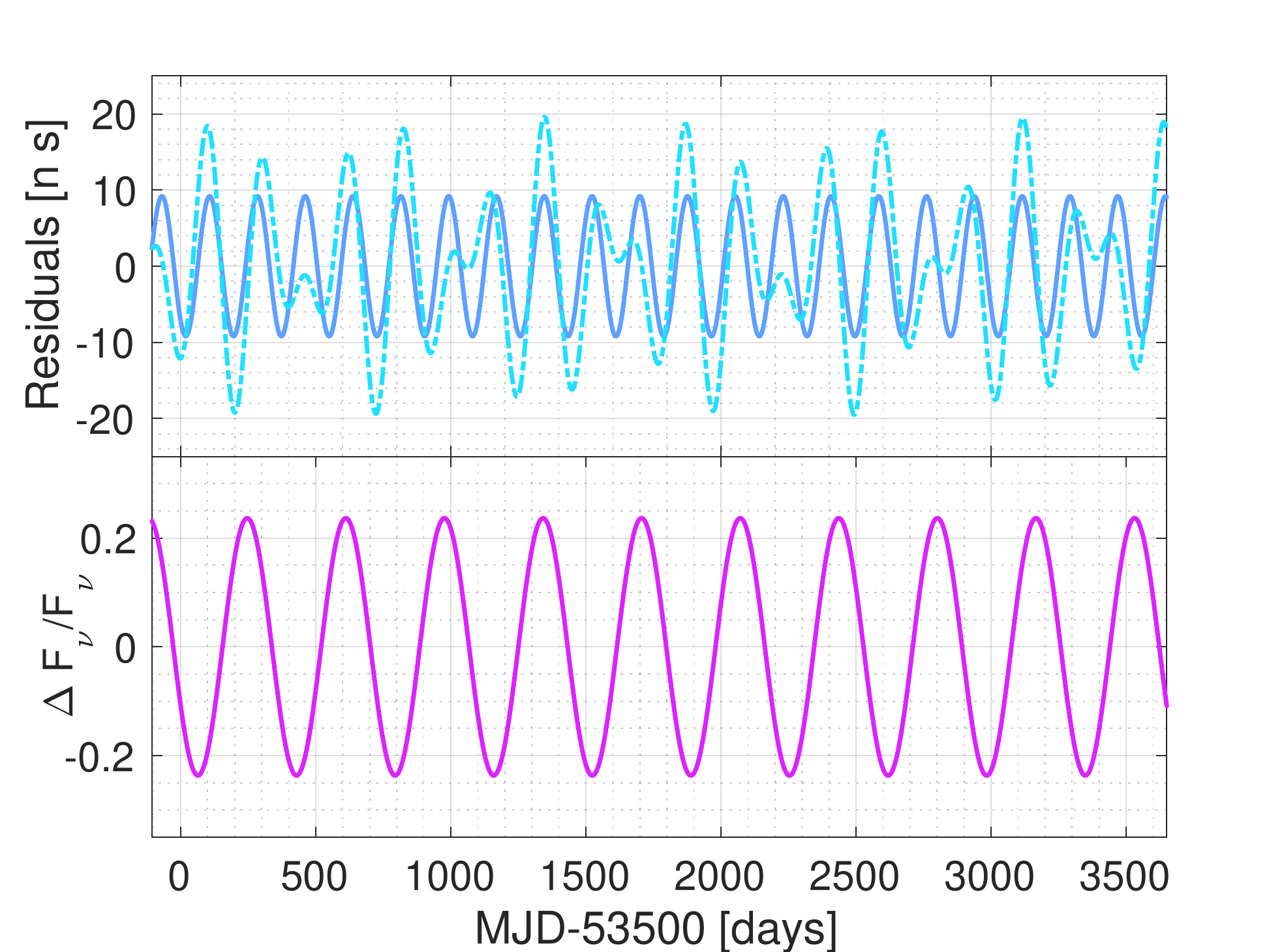}
\caption{Simulated GWs and electromagnetic signatures for a binary with the fiducial parameters. The top panel shows the timing residuals this binary produces from the Earth-term (solid dark blue line) and from both the Earth- and pulsar-term, i.e., the full signal (dashed light blue line). The bottom panel shows the corresponding Doppler boost signature.} 
\label{fig:Residuals_circular}
\end{figure}

\begin{figure} 
    \centering
     \includegraphics[width=\columnwidth]{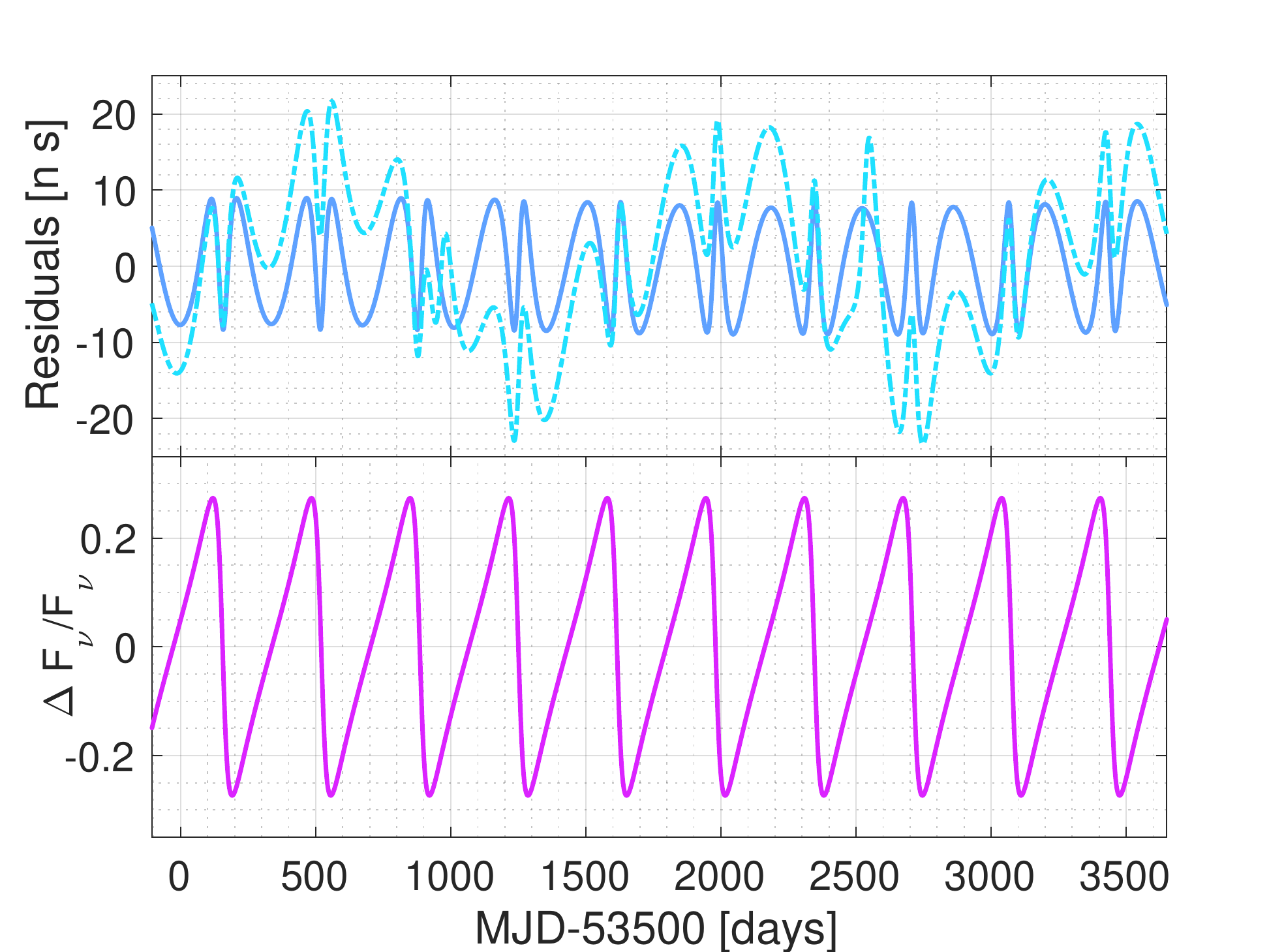}
\caption{Similar to \autoref{fig:Residuals_circular}, but for a binary with eccentricity $e=0.5$.} 
\label{fig:Residuals_eccentric}
\end{figure}

In \autoref{fig:Residuals_circular} and \autoref{fig:Residuals_eccentric}, we show two examples of how GW and electromagnetic signals can be correlated for a circular and eccentric binary, respectively.
In the top panel of each figure, we show simulated timing residuals that a fiducial binary induces on a pulsar with an observing baseline of 10\,yr. The binary is placed at a distance $D_L=20$\,Mpc (i.e. comparable to the distance to the Virgo cluster). We present the Earth-term residuals with a solid dark blue line and the full signal which includes the pulsar-term with a dashed light blue line. In the bottom panel, we show the corresponding Doppler boost signal calculated from \autoref{eq:DB_ciruclar}. For the eccentric binary, the eccentricity is set to $e=0.5$.

Next we explore for which binaries joint detections are possible, given current and future sensitivities of PTAs. We consider a Doppler boost signal to be detectable if the peak-to-peak amplitude is at least 10\%. From \autoref{eq:DB_ciruclar}, we see that the amplitude depends on the spectral index, which we fix at the fiducial value $\alpha_{\nu}=-0.44$ from the composite quasar spectrum \citep{2001AJ....122..549V}. In addition, the amplitude depends on the line-of-sight velocity of the secondary, and in turn on the binary total mass, mass ratio, period and inclination, each of which we sample over a representative parameter space. For the GW signal, we compare the strain of the binary with current upper limits from the NANOGrav 11\,yr dataset \citep{NANOGrav_CWs_2018}.\footnote{The 11\,yr dataset provides the most up-to-date upper limits on continuous waves, but the subsequent 12.5yr dataset has also become available \citep{2021ApJS..252....4A}.} We also use projected future sensitivities from \citet{2021ApJ...915...97X} for an IPTA dataset in 2025 and a PTA sensitivity in 2034 enhanced due to a significant increase in the number of pulsars identified by the Square Kilometer Array (SKA).
We use sky-averaged sensitivity curves, but we note that PTA sensitivities are highly heterogeneous and vary by almost an order of magnitude from the most sensitive to the least sensitive part of the sky.

\begin{figure}    
\begin{subfigure}{\columnwidth}
    \centering
 \includegraphics[width=\columnwidth]{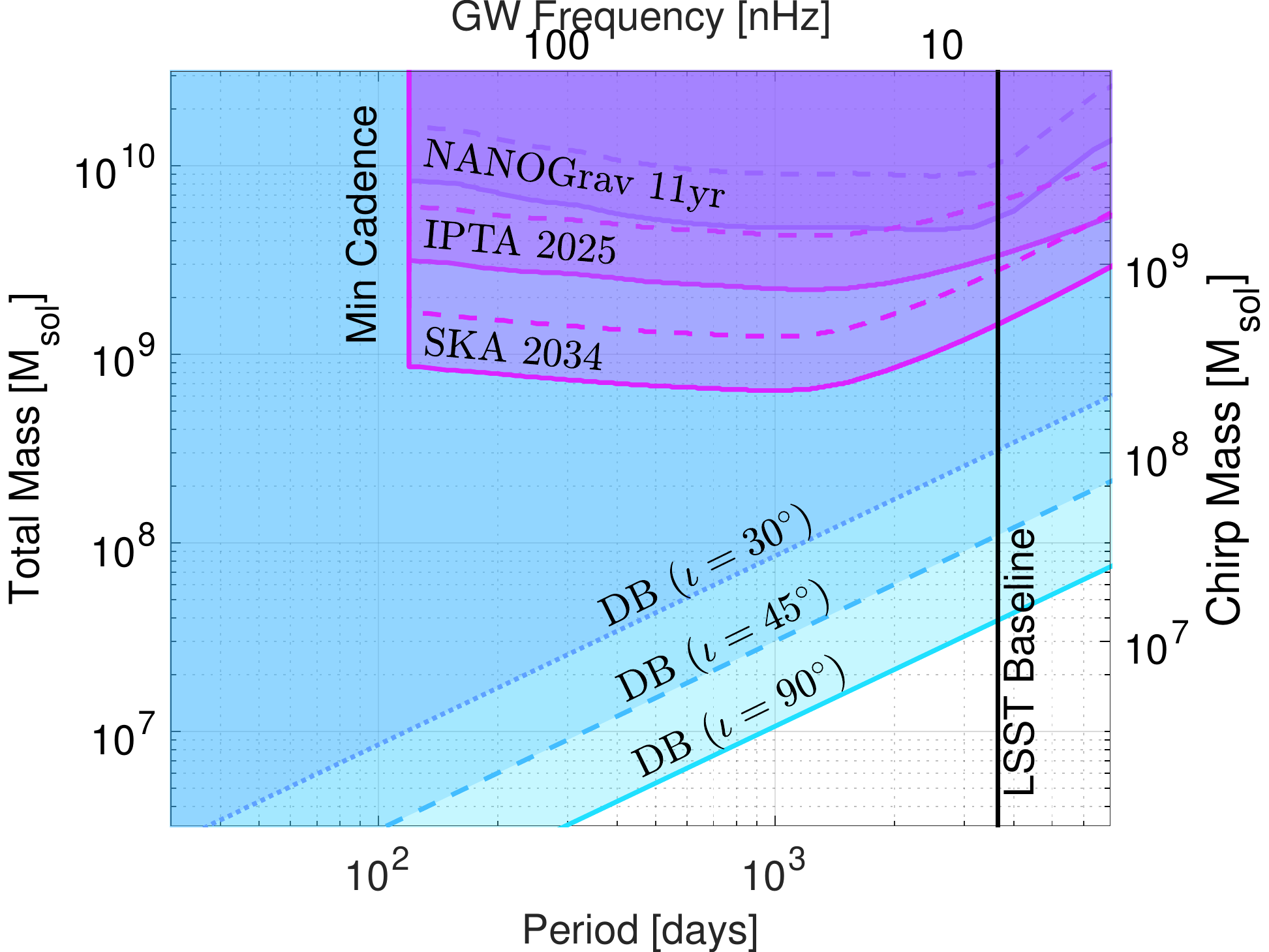}
 \end{subfigure}
 
 \begin{subfigure}{\columnwidth}
      \includegraphics[width=\columnwidth]{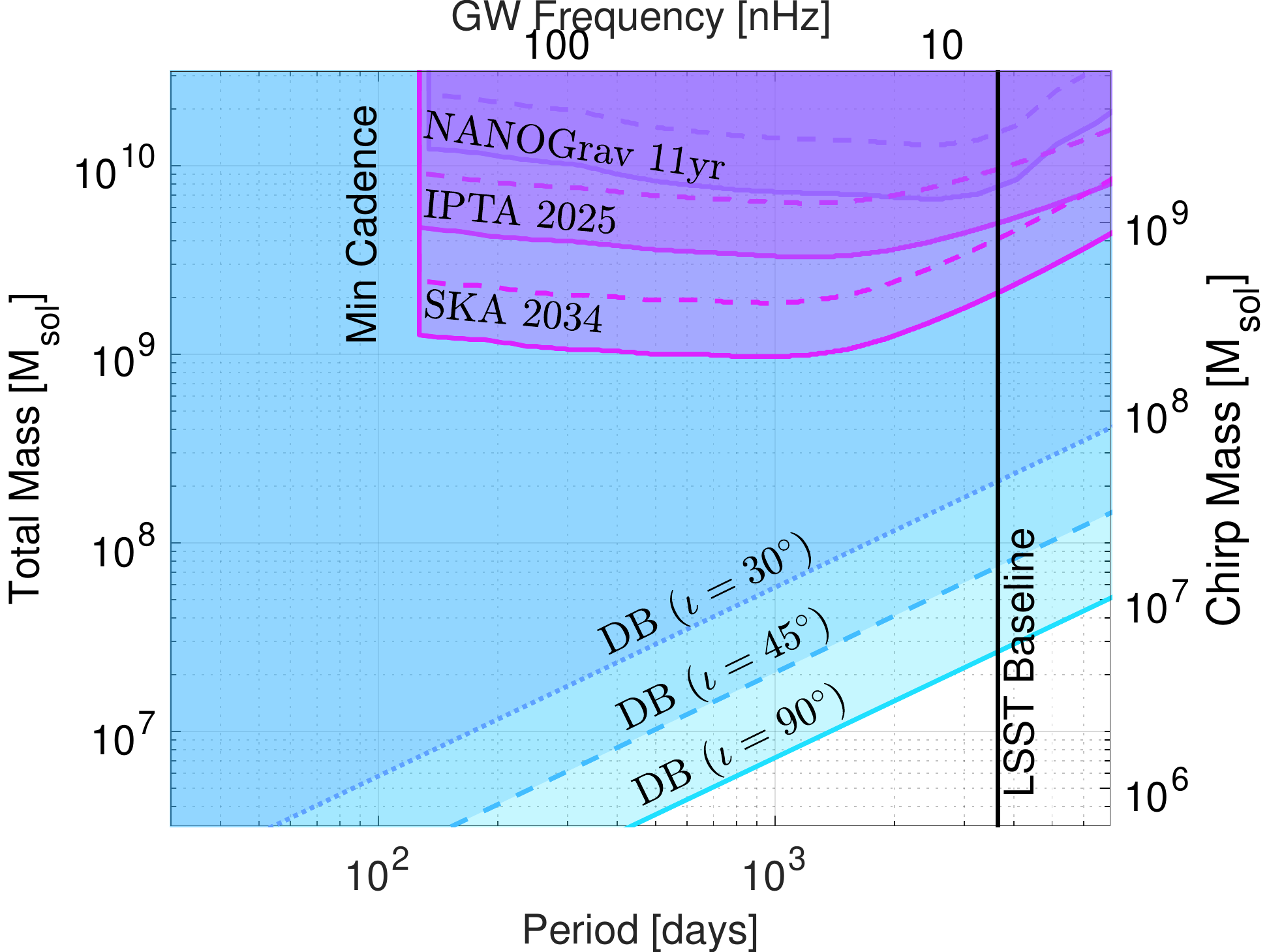}
 \end{subfigure}
 
 \caption{Binary parameter space (total/chirp mass vs binary period/GW frequency) of detectable Doppler boost and GW signals for binaries with $q=0.25$ (top panel) and $q=0.01$ (bottom panel). With the blue shaded region we highlight Doppler boost signatures with peak-to-peak amplitudes $\geq$10\% for varying inclinations, and with the purple shaded region we show binaries detectable by PTAs based on current and future sensitivities, with the solid and dashed lines indicating binaries at 100\,Mpc and 300\,Mpc.}
 \label{Fig:param_space}
\end{figure}

In \autoref{Fig:param_space}, we delineate the parameter space of binaries as a function of total mass (or chirp mass) and orbital period (or GW frequency) that can produce bright electromagnetic signals and strong GWs. With blue shaded regions, we show the detectable Doppler signals for circular binaries for fixed mass ratios ($q=0.25$ for the top panel and $q=0.1$ for the bottom panel) and varying inclinations. Binaries on edge-on orbits ($i=90^{\circ}$) are detectable for a wider range of parameters highlighted with light blue shade compared to binaries closer to edge-on (e.g., $i=30^{\circ}$, shaded with a darker blue color). We overplot with the purple shaded regions the binaries detectable in GWs by PTAs given the current sensitivity from the NANOGrav 11\,yr dataset, and projected future sensitivities from IPTA in 2025 and SKA in 2034. The solid and dashed lines are for binaries at 100\,Mpc and 300\,Mpc, respectively. 

We see that for the current PTA sensitivity, the common parameter space is rather limited and multi-messenger observations are possible only for the most massive and nearby galaxies. However, as the PTA sensitivity increases in the next 5-10\,yr, more and more binaries will satisfy the criteria for detection both in GWs and in electromagnetic time-domain surveys. From \autoref{Fig:param_space}, we also see that the parameter space for joint multi-messenger observations depends on the mass ratio. For instance, the Doppler boost signature is more prominent for lower mass binaries (and thus the shaded blue region is more extended in the case of $q=0.1$), whereas GW emission is stronger for high mass ratio binaries (and thus the purple shaded region is larger for $q=0.25$). For higher mass binaries, PTAs can cover more binaries, but the Doppler boost emission becomes less dominant for these mass ratios. We expect similar trends for Doppler boost emission from eccentric binaries, but since PTAs currently place limits only on binaries in circular orbits, we cannot derive the respective common parameter space. Future work should address this limitation.

Beyond Doppler boost variability, connecting the GW signal with periodic variability arising from periodic accretion is more challenging. For unequal-mass binaries ($q<0.3-0.5$), the variability period coincides with the orbital period and can be linked to the GW frequency as $f_{\rm GW}=2/P$. The connection is weaker, because the amplitude and the phase cannot be linked to the orbital parameters. The common parameter space is similar to \autoref{Fig:param_space}, with the shaded purple area for GWs being identical. However, the detectability of such electromagnetic signals depends on their amplitude, which is uncertain. Likely the most massive binaries produce the brightest and most detectable signals, which is very promising for multi-messenger detections.
For higher mass ratios, the most prominent periodic component occurs at a few (5-8) times the orbital period, which makes the link with the GW signal even more tentative. Our ability to perform multi-messenger analysis with such signals may rely on the additional periodic components (at the orbital and half the orbital period of the binary), especially if those are due to relativistic Doppler boost.
Eccentric binaries may provide an exception to this. Simulations of eccentric equal mass binaries predict periodic bursts repeating at the orbital period of the binary with the peak of the burst occuring within 10-20\% of the pericenter passage \citep{2016ApJ...827...43M,2017MNRAS.466.1170M,2021ApJ...909L..13Z}.

Last but not least, multi-messenger observations based on Doppler shifted emission lines are even more challenging. This method probes binaries with longer periods of at least a few decades and thus the overlap with PTAs is currently limited. However, a few of the best monitored pulsars have timing observations spanning over two decades, and thus can probe binaries with orbital periods up to 40\,yrs. As additional data are collected, the low frequency limit will be expanded. In the meantime, additional observations of existing SMBHB candidates will also expand the radial velocity curves leading to better constraints on the binary parameters. Therefore, even if this signature is currently less promising, within 10\,yrs there could be significant potential for multi-messenger observations.

\section{Discussion}
\label{sec:discussion}

We have presented GW signals of SMBHBs along with their respective electromagnetic signatures. We demonstrated that for a variety of binaries the periodic signals deviate from simple sinusoidal behavior (e.g., eccentric binaries, periodic accretion). This paper provides the analytic framework for a broad variety of binary signals that can serve as templates for future searches for non-sinusoidal periodicity. Since current electromagnetic searches for quasars with periodic variability are optimised for (and thus limited to) quasi-sinusoidal signals, these templates are of great significance.

In this section, we consider a few additional instances, in which the binary emission likely departs from purely sinusoidal. For instance, we relax some of the basic assumptions in the simplest Doppler boost model presented in \S~\ref{sec:DB_simplest}. We also explore potential deviations from the fact that photometric magnitudes employed in optical time-domain surveys depend logarithmically on the binary flux.

\subsection{Circular binary with variable mini-disk luminosity}
\label{sec:DB_variable_lum}
In the simplest case of Doppler boost emission (presented in \S~\ref{sec:doppler_boost}), we assumed that the emission of the secondary mini-disk $F_{\nu,0}^{\rm sec}$ (which is the only one contributing to the variability) is constant. This is likely the case for very unequal binaries (e.g., $q<0.05$), for which the accretion is found to be steady \citep{2013MNRAS.436.2997D}.
However, for higher-mass ratios, simulations suggest a more dynamic picture. For instance, streams enter the cavity and hit the mini-disks, potentially leading to instantaneous or even long-lasting changes in the mini-disk luminosity. The two-mini disks may also interact, leading to the flow of material from one to the other, which can also alter the mini-disk luminosity that gets Doppler boosted \citep{Bowen_2017b}. 

Currently there is no analytical prescription for potential variability in mini-disk emission. Hence we introduce a simple model to capture the above effect. Specifically, we add stochastic fluctuations $\delta F(t)$ to the luminosity of the mini-disk drawn from a Gaussian distribution $\mathcal{N}(\mu,\sigma^2)$ with zero mean ($\mu=0$) and standard deviation equal to a fraction $f$ of the constant stationary luminosity of the secondary mini-disk ($\sigma= f\times F_{\nu,0}^{\rm sec}$), with $0\leq f\leq 1$. Therefore
\begin{align}\label{eq:lum_sec_variable}
    F_{\nu}^{\rm sec}(t) &= F_{\nu,0}^{\rm sec}+\delta F(t)=F_{\nu,0}^{\rm sec}+\mathcal{N}(0,(f\times F_{\nu,0}^{\rm sec})^2)\nonumber\\&
    =F_{\nu,0}^{\rm sec}\left[1+\mathcal{N}(0,f^2)\right].
\end{align}

\begin{figure} 
 \includegraphics[width=\columnwidth]{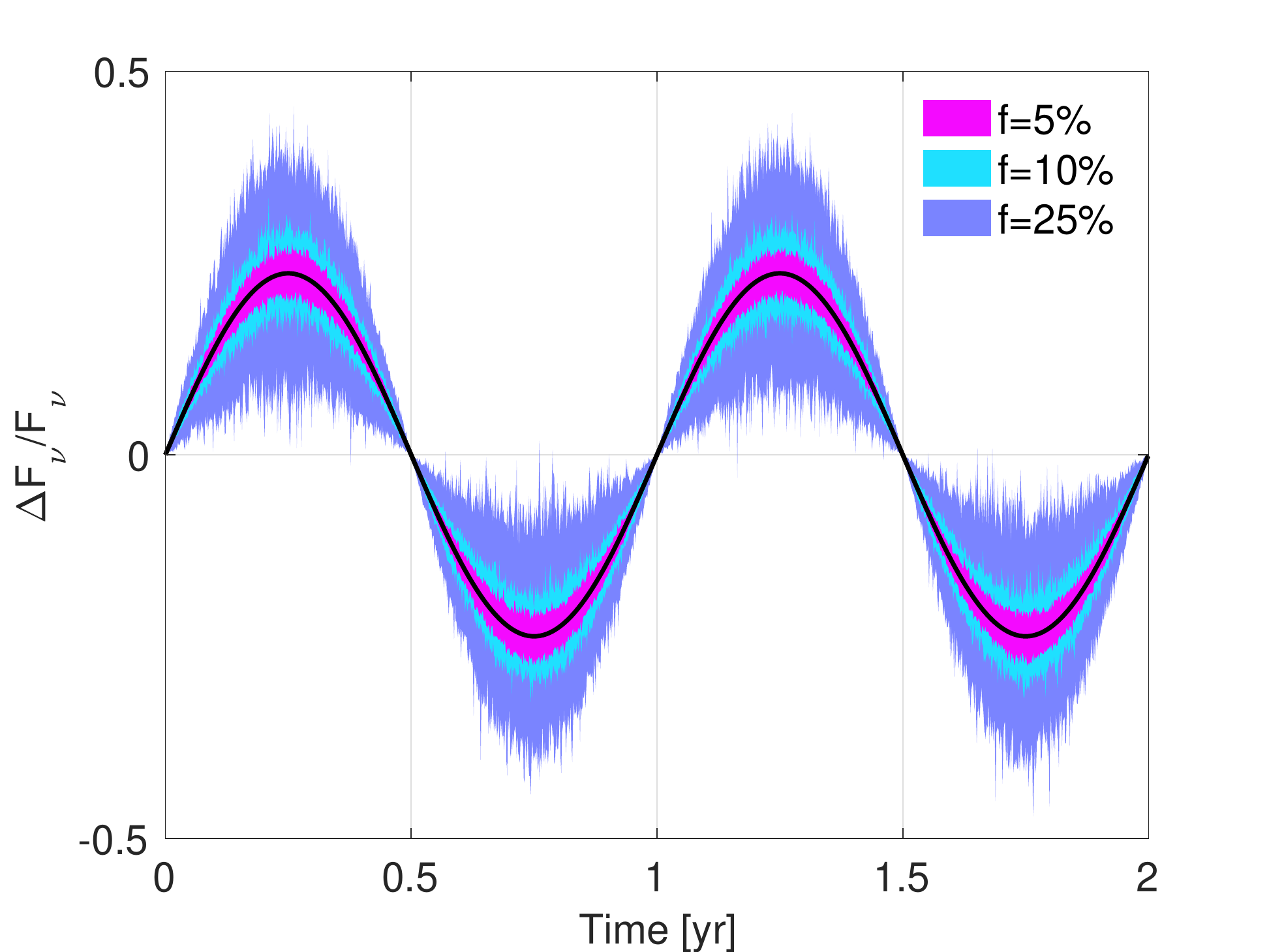}
\caption{Doppler boost variability from 100 realizations of the fiducial binary, in which the luminosity of the secondary mini-disk varies over time, with fluctuations at levels from 5\% (magenta), 10\% (light blue) and 25\% (dark blue) of the constant luminosity.} 
\label{fig:variable_secondary}
\end{figure}

In \autoref{fig:variable_secondary} we illustrate the Doppler boost variability resulting from variable luminosity of the secondary mini-disk, as prescribed by \autoref{eq:lum_sec_variable}. We present the result of 100 realizations, varying $f$ from 5\% (magenta), 10\% (light blue) and 25\% (dark blue), for our fiducial binary ($M_{\rm tot}=10^9M_{\odot}$, mass ratio $q=0.25$ and period $P=1$\,yr and $i=60^{\circ}$). For the spectral slope, we adopt the value $\alpha_{\nu}=-0.44$ from the composite quasar spectrum in \citet{2001AJ....122..549V}. 
It is apparent that fluctuations that are temporally close to the maximum/minimum velocity (e.g., points C and G in \autoref{fig:V_LOS}) will be amplified significantly, causing departures from the overall sinusoidal behavior of the light curve. By contrast, fluctuations occurring when the orbital velocities are perpendicular to the line-of-sight (e.g., points A and E) will have zero impact. This is expected, since the Doppler variability depends on the L.O.S. velocity (e.g., \autoref{eq:DB_ciruclar}), but it does have potential implications. 

First, significant deviations from sinusoidal variability may render such binaries undetectable. The detectability of such variability likely depends on the interplay between the cadence (i.e. frequency) of observations and the frequency and amplitude of fluctuations. If the cadence is dense, and enough data points are collected within a cycle, the sinusoidal variability can be recovered as a median curve of the flux distribution, especially if many cycles are observed. If observations are sparse, then frequent instantaneous deviations would in general preclude the detection of periodicity (see also \citealt{2021arXiv211007465W}). Additionally, since the minima and maxima are amplified, it may lead to biased constraints of the variability amplitude and thus of the binary parameters. 

Last, but not least, the perturbations we introduced above are a simple version of instantaneous fluctuations in the secondary mini-disk luminosity. 
In reality, shock heating from gaseous streams pulled from the cavity, or gas overflow from the secondary to the primary mini-disk, may lead to significant increases/decreases in the luminosity of the secondary mini-disk (and by extension to the variability amplitude) that last for a significant fraction of the orbit. 
It is worth noting that one of the most prominent binary candidates, quasar PG$1302$$-$$102$, shows a slightly increased amplitude in its periodicity in the last observed cycle. This feature was originally suggested to disfavor the Doppler boost model \citep{2018ApJ...859L..12L}. However our results here indicate that such deviations can be produced within the Doppler boost model if the mini-disk luminosity is not constant. Hence, further analysis of this candidate is required.

The fluctuations described above likely occur on top of the typical AGN variability, best described by a Damped Random Walk (DRW) model \citep{2009ApJ...698..895K,2010ApJ...721.1014M}. In the case of a binary, DRW variability could be associated with the circumbinary disk or the accretion processes in the mini-disks (or both).
In fact, the DRW model offers an alternative option to simulate the stochastic variations of the secondary mini-disk luminosity. We defer this analysis for a future study, in which we will include realistic noise modeling. However, we note that the DRW is also unlikely to capture the fluctuations in luminosity caused by the dynamic interplay between the two mini-disks and the circumbinary disk.

\subsection{Circular binary with bright primary mini-disk}
So far we have assumed that
only the typically more luminous and faster moving secondary mini-disk contributes to Doppler boost variability (\S~\ref{sec:doppler_boost} and \S~\ref{sec:DB_variable_lum}). In highly unequal-mass binaries, this is a sensible choice, but as the mass ratio increases, the luminosity of the primary mini-disk may also become significant. Considering the contribution of the primary mini-disk, the overall variability is

\begin{align}
\label{eq:DB_ciruclar_primary}
    F_{\nu}&=F_{\nu,0}^{\rm sec}[1+(3-\alpha_{\nu})v_2/c\sin(i)\sin(\omega t+\phi_0)]\nonumber\\  
    &+F_{\nu,0}^{\rm prim}[1+(3-\alpha_{\nu})v_1/c\sin(i)\sin(\omega t+\pi+\phi_0)]
\end{align}

\begin{figure}
 \includegraphics[width=\columnwidth]{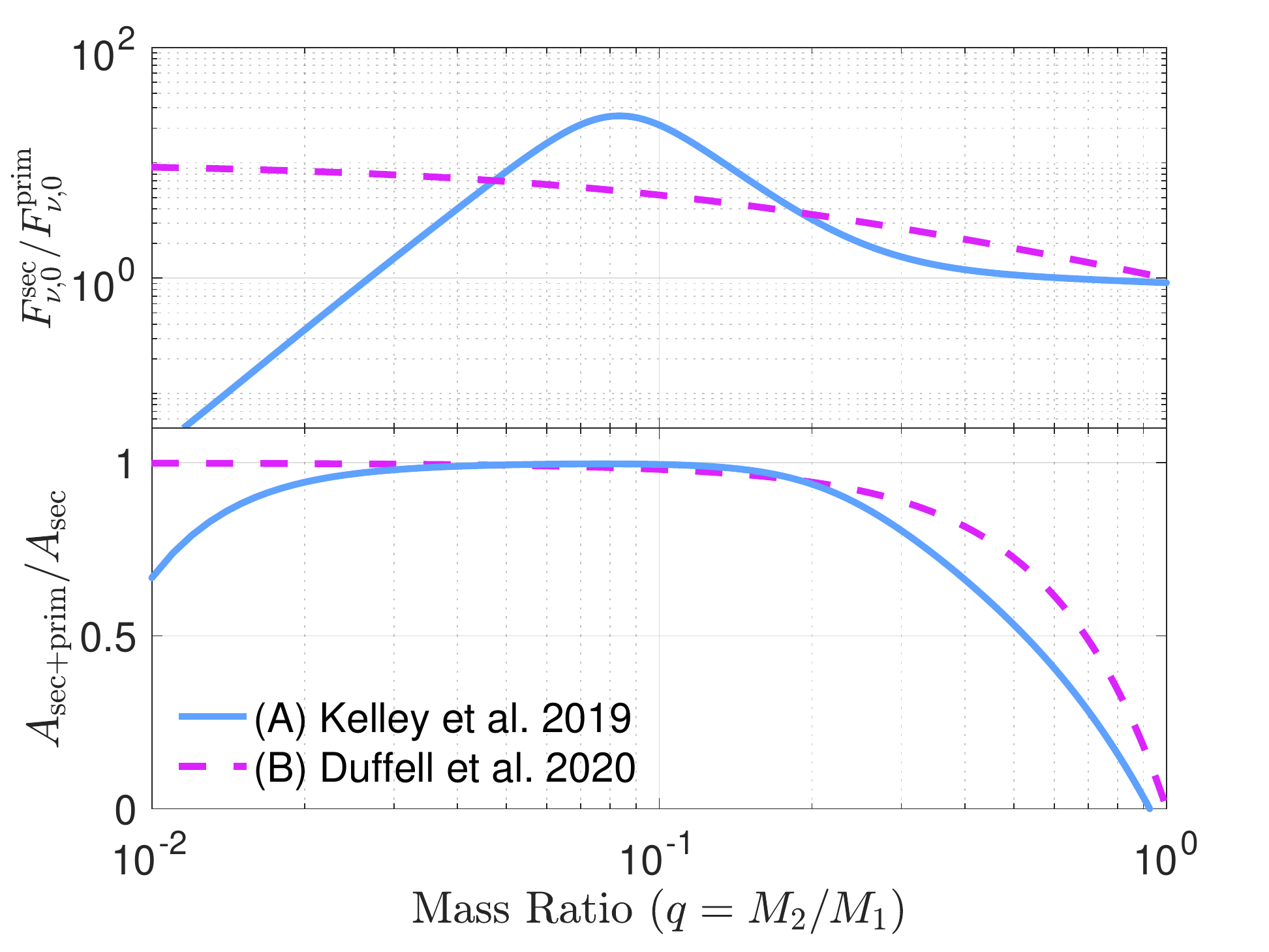}
 \caption{\emph{Top Panel}: Comparison of the luminosity ratio (assumed to be equal to the relative mass accretion rate) as a function of mass ratio from (A) \citet{2019MNRAS.485.1579K}, and (B) \citet{2020ApJ...901...25D}. \emph{Bottom panel}: Relative amplitude of the Doppler boost variability when both mini-disks are included over the respective amplitude when only the secondary mini-disk is included versus mass ratio.}
 \label{Fig:primary_included}
\end{figure}

Since the two SMBHs are moving in opposite directions, the motion of one will lead to an increase of the luminosity, while, at the same time, the motion of the other will have a dimming effect. In order to quantify the aggregate effect, we need the relative luminosity of the two mini-disks. Assuming that these are proportional to the accretion rate onto each SMBH, we adopt two approaches from the literature, which we present in the top panel of Fig.~\ref{Fig:primary_included}. As option (A), we use the fitting formula from \citet{2019MNRAS.485.1579K} based on the simulations of \citet{2014ApJ...783..134F}.
\begin{equation}
   \frac{F_{\nu,0}^{\rm sec}}{F_{\nu,0}^{\rm prim}}\approx\frac{\dot{M}_2}{\dot{M}_1} = q^{-1/4}e^{-1/10q} + \frac{50}{(12q)^{7/2} + (12q)^{-7/2}}
\end{equation}
As an alternative option (B), we employ a simpler analytical form from the more recent work of \citet{2020ApJ...901...25D}.
\begin{equation}
    \frac{F_{\nu,0}^{\rm sec}}{F_{\nu,0}^{\rm prim}}\approx\frac{\dot{M}_2}{\dot{M}_1} = \frac{1}{0.1+0.9q}
\end{equation}

In the bottom panel of Fig.~\ref{Fig:primary_included}, we present the relative amplitude of the Doppler boost variability when both SMBHs are included versus the amplitude when only the luminosity of the secondary mini-disk is considered as a function of the mass ratio. For an extensive range of mass ratios ($0.03<q<0.2$), the inclusion of the primary has practically no effect, which justifies the choice of previous studies to completely ignore its contribution. However, as the mass ratio increases, the contribution of the primary becomes more important. For instance, the amplitude gradually decreases to 50\% at $q=0.5$ or $q=0.7$ for prescription (A) and (B), respectively. As the mass ratio increases further, the amplitude of Doppler boost variability diminishes. 

Our results are in agreement with previous studies suggesting the Doppler boost mechanism is sub-dominant (e.g., compared to other sources of periodicity like periodic accretion) for binaries approaching equal-mass \citep{Bowen_2017b,Tang+2018}. On the other end of the mass ratio distribution ($q<0.03$), our results deviate depending on the adopted relative luminosity on the two-mini disks; for prescription (A) the amplitude decreases significantly, whereas for (B) the contribution of the primary is negligible. We note, however, that in the limit of low mass ratio ($q\rightarrow 0$), the flux ratio also diminishes ($F_{\nu,0}^{\rm sec}/F_{\nu,0}^{\rm prim}\rightarrow 0$), and thus option (A) may be more appropriate for low mass ratios. For the purpose of multi-messenger observations, this regime is slightly less promising at least in the near-future. The strain of the GW signal depends on the mass-ratio, and given the current sensitivities of PTAs, highly unequal binaries are unlikely to be the first to be detected (see \S~\ref{sec:mma}).

\subsection{Binary signatures in photometric magnitudes}
We have presented our analysis of the Doppler boost variability in terms of flux. However, optical and UV data are typically expressed in magnitudes rather than fluxes. For this, we convert \autoref{eq:DB_final} in magnitudes. 
The apparent magnitude $m_{\nu}$ of a source in a narrow-band filter around frequency $\nu$ is related to the observed flux as
\begin{equation}
    m_{\nu}=-2.5\log_{10}\left(F_{\nu}\right)+m_{\rm ref}
\end{equation}
where $m_{\rm ref}=-2.5\log_{10}\left(F_{\rm ref}\right)$ is the  reference magnitude for the flux $F_{\rm ref}$ in the specific photometric filter. Then, from \autoref{eq:DB_final}

\begin{align}
    \label{eq:mag_from_flux}
    m_{\nu}&=-2.5\log_{10}\left(F_{\nu,0}\left(1+\Delta F\right)\right)+m_{\rm ref}\nonumber\\
    &=-2.5\log_{10}\left(F_{\nu,0}\right)-2.5\log_{10}\left(1+\Delta F\right)+m_{\rm ref}\nonumber\\
    &=m_{\nu,0}-\frac{2.5}{\ln(10)}\ln\left(1+\Delta F\right)
\end{align}
For small $\Delta F$ ($\Delta F<<1$), the first order approximation gives $\ln\left(1+\Delta F\right)\approx\Delta F$ which leads to
\begin{equation}
\label{eq:mag_flux_aprox}
    m_{\nu}\approx m_{\nu,0}-1.08 \Delta F
\end{equation}

The above approximation provides a linear correlation between magnitude and flux, which can easily translate the Doppler boost variability in magnitudes. As expected the bright phase of Doppler boost ($\Delta F>0$) leads to a decrease in magnitude, whereas the dimming phase ($\Delta F<0$) leads to magnitude increase. Previous studies \citep{Dorazio2015Nature,2018MNRAS.476.4617C,2020MNRAS.496.1683X} have taken advantage of this proportionality, in order to fit the Doppler boost model directly in observed optical/UV light curves.

\begin{figure} 
 \includegraphics[width=\columnwidth]{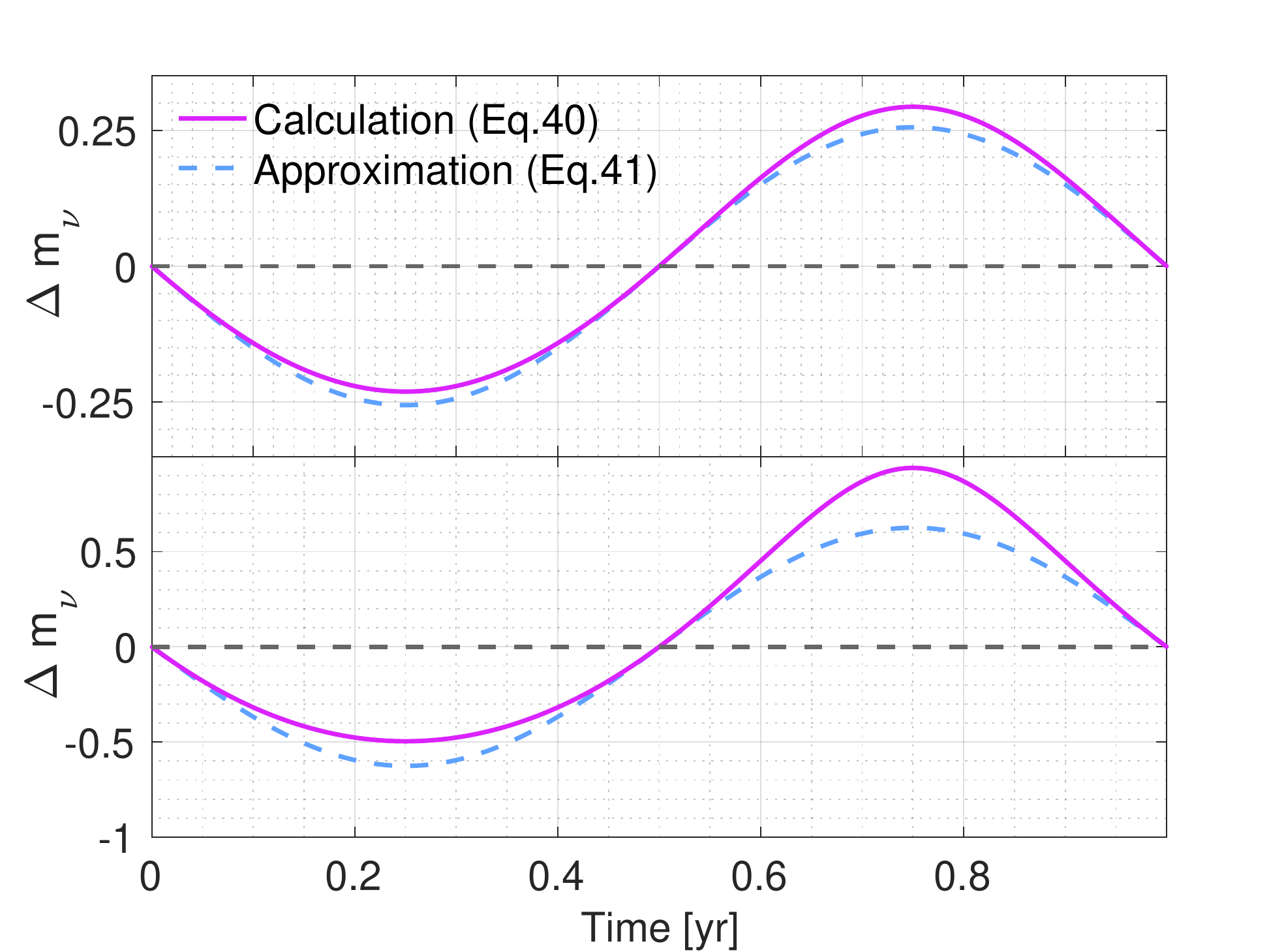}
\caption{Doppler boost light curve in magnitudes for a binary with the fiducial parameters (top panel), and for a binary with $M_{\rm tot}=10^{10}M_\odot$ and $q=0.1$ (bottom panel). The dashed blue line shows the magnitude estimated directly from the flux \autoref{eq:mag_from_flux}, whereas the solid purple line shows the approximate magnitude estimate from \autoref{eq:mag_flux_aprox}.} 
\label{fig:magnitude}
\end{figure}

In \autoref{fig:magnitude}, we show the Doppler boost variability expressed in magnitudes calculated directly from the flux (\autoref{eq:mag_from_flux}) with solid purple line, and estimated from the approximation in \autoref{eq:mag_flux_aprox} with dashed blue line. We present two cases, a binary with the fiducial parameters ($M_{\rm tot}=10^9M_{\odot}$, mass ratio $q=0.25$ and period $P=1$\,yr and $i=60^{\circ}$, $\alpha_{\nu}=-0.44$) on the top panel and a binary with $M_{\rm tot}=10^{10}M_\odot$ and $q=0.1$ on the bottom panel. The magnitude approximation results in sinusoidal and symmetric variability (since it follows exactly the flux variations), whereas it becomes more asymmetric and deviates from the sinudoid with the more direct calculation. This effect is more pronounced for higher Doppler boost amplitudes and is diminished when the Doppler boost amplitude is small. For instance, in the above example, the deviation is $\sim9\%$ and $\sim15\%$ at the minimum and maximum of the sinusoid for the fiducial binary, but $\sim40\%$ and $\sim50\%$ for the more massive and unequal-mass binary. Therefore, even though using light curves as provided by the surveys (i.e. in magnitudes) is convenient, based on the above, we recommend the use of fluxes for the joint multi-messenger analysis.

\section{Future Prospects}
\label{sec:summary}

The next ten years promise to revolutionize the field of SMBHBs. PTAs may be on the cusp of detecting a stochastic GW background. The first sign of this detection would be through the emergence of an excess low-frequency noise process that has a common spectrum across the monitored pulsars \citep{2021PhRvD.103f3027R}. Indeed, NANOGrav has recently seen such a low-frequency common process in $45$ pulsars from its $12.5$-year baseline dataset \citep{2020ApJ...905L..34A}. This has recently been validated by the PPTA \citep{2021arXiv210712112G}.
Definitive detection will require the measurement of distinctive GW-induced quadrupolar-correlated timing deviations between the spatially-separated pulsars \citep{Hellings1983}, which could occur within the next couple of years 
\citep{2021ApJ...911L..34P}. 

PTAs continue their march towards ever greater sensitivity through lengthened baselines of existing pulsars and adding newly discovered pulsars to the monitored arrays. Current survey programs add $\sim 4-6$ pulsars per year to monitored arrays, which are highly beneficial to GW background probes and the prospects for localizing individual SMBHB sources. The sensitivity can also be increased through the combination of distinct datasets into an IPTA dataset, which also has the advantage of more uniform coverage of the sky (e.g., \citealt{Verbiest2016,pdd+19}). Based on these improvements theoretical studies forecast that several individual binaries may be resolved by PTAs by the end of the 2020s \citep{Rosado2015,Kelley2018,Mingarelli2017}. 

On the electromagnetic side, spectroscopic and photometric  datasets of unprecedented quality and quantity like SDSS~V, and the Vera C. Rubin Legacy Survey of Space and Time (LSST) will uncover larger numbers of candidates. SDSS~V \citep{Kollmeier2017} is an ongoing time-domain spectroscopic survey that will provide repeat observations of over $10^5$ quasars. In addition to the unprecedented sample size, the observations will span longer timescales than previous samples, which is important given that SMBHB periods found with spectroscopy are expected to be long. SDSS~V will provide many new spectroscopic candidates in the spirit of \citet{eracleous12} and define the benchmark spectroscopic variability properties of regular quasars that are needed to refine such candidates.

LSST will detect between 20 and 100 million quasars \citep{2009arXiv0912.0201L,2021MNRAS.506.2408X} and provide well-sampled light curves in multiple bands for a baseline of 10 years. Based on these, theoretical models have predicted that in LSST several hundreds of sub-parsec SMBHBs should be detectable as quasars with periodic variability \citep{2019MNRAS.485.1579K,2021MNRAS.506.2408X,2021arXiv210707522K}. Another exciting possibility is that for a subset of quasars, their baselines can be extended with already existing light curves. We note that even though most current time-domain surveys cover the northern hemisphere (with the exception of CRTS which is an all-sky survey), while LSST is a southern hemishpere survey, we expect significant overlap around the equatorial plane. These long baselines are crucial for SMBHB searches, because they allow us to observe several cycles of periodicity even if the periods are relatively long of order a few years. 

The above advancements will lead to the first multi-messenger detection of SMBHBs. This paper prepares the way for the combination of electromagnetic time-domain data and GW data from PTAs. In a future companion study, we will simulate a population of SMBHBs with realistic electromagnetic and GW signatures. Employing realistic noise properties both for light curves and for an array of pulsars, we will perform a joint multi-messenger analysis to assess the advantages (improved parameter estimation, boost in detectability, etc.) of a combined detection. We will develop a multi-messenger pipeline that will become publicly available for community use.

This paper provides the mathematical foundation for a multi-messenger analysis of electromagnetic and GW signals of SMBHBs. In particular here, we connected the signatures to the orbital dynamics and explored the connection between the two signals. Our work is summarized as follows:
\begin{itemize}
    \item We described the orbital dynamics of circular and eccentric binaries.
    \item We presented the GW residuals that an eccentric binary induces to a PTA and we connected the GW strain to the binary orbit.
    \item We described promising electromagnetic signatures of SMBHBs (Doppler shifted emission lines, periodic variability due to relativistic Doppler boost, periodic variability due to periodic accretion). We demonstrated how each signal is connected to the orbital elements of the binary.
    \item We presented the common temporal evolution of timing residuals and light curves for tentative binaries.
    \item We showed that Doppler boost and PTAs have significant overlap in the binary parameter space (period, total mass) for current and projected future sensitivities. 
    \item Currently (based on the 11\,yr dataset of NANOGrav) only the most massive binaries can be covered with both methods ($M_{\rm tot}\simeq 10^{10}M_{\odot}$ for distances up to 300\,Mpc and mass ratios $q=1/4$), but as the sensitivity increases the common parameter space is expected to significantly expand in the next 5-10yrs.
    \item We demonstrated that deviations from the simplest Doppler boost model (either due to variable luminosity in the secondary mini-disk or from the inclusion of the primary mini-disk luminosity) can introduce deviations from sinusoidal variability or reduce its expected amplitude.
    \item We showed that it is preferable to use light curves expressed in fluxes (instead of magnitudes)  for the joint multi-messenger analysis.
    \item A companion study will include a publicly available pipeline that combines light curves and PTA data.
\end{itemize}

\section*{Acknowledgements}
We thank our colleagues in NANOGrav and the International Pulsar Timing Array for useful discussions about this work. Special thanks to Sarah Vigeland and Jeff Hazboun for providing the NANOGrav and IPTA/SKA sensitivity curves, and to Luke Kelley and Dan D'Orazio for fruitful discussions during the preparation of the manuscript. MC, SRT, JR, TB and JRT acknowledge NSF AST-200793. SRT acknowledges support from NSF PHY-2020265. SRT and JR also acknowledges support from Vanderbilt University College of Arts \& Science Dean's Faculty Fellowship program. T.B. acknowledges the support from the NASA award 80NSSC19K0319 and from the NSF award AST-1908042. JRT acknowledges support from NSF grants CAREER-1945546, AST-2009539, and AST-2108668.



\bibliographystyle{mnras}
\bibliography{mma} 






\bsp	
\label{lastpage}
\end{document}